\documentclass{mn2e}
\usepackage{graphics}
\usepackage{epsfig}

\def\ni{\noindent}                                       %No indent%
\def\ls{\vskip 12.045pt}                            %One Line space%
\def\etal{et\thinspace al.\ }                               %et al.%
\def\etp{et\thinspace al.}    %et al., but with no space at the end%

\def\ojo{\fbox{\bf !$\odot$j$\odot$!}}      %Olho! Needs Correction%
\title[Stellar Populations of Low Luminosity Active Galactic]{The
      Stellar Populations of Low Luminosity Active Galactic
      Nuclei. III: Spatially Resolved Spectral Properties}

\author[Cid Fernandes \etal]
       {R. Cid Fernandes$^{1}$\thanks{E-mail: cid@astro.ufsc.br}, 
	 R. M. Gonz\'alez Delgado$^{2}$\thanks{E-mail: rosa@iaa.es},
	 T. Storchi-Bergmann$^{3}$\thanks{E-mail: thaisa@if.ufrgs.br},
	 \newauthor
	 L. Pires Martins$^{4}$\thanks{E-mail: martins@stsci.edu},
	 H. Schmitt$^{5}$\thanks{E-mail: hschmitt@nrao.edu. Jansky Fellow.}\\
	 $^{1}$ Depto.\ de F\'{\i}sica - CFM - Universidade Federal de
	 Santa Catarina, C.P. 476, 88040-900, Florian\'opolis, SC,
	 Brazil\\
	 $^{2}$ Instituto de Astrof\'{\i}sica de Andaluc\'{\i}a (CSIC),
	 P.O. Box 3004, 18080 Granada, Spain\\
	 $^{3}$ Instituto de F\'{\i}sica, Universidade Federal do Rio
	 Grande do Sul, C.P. 15001, 91501-970, Porto Alegre, RS, Brazil\\
	 $^{4}$ Space Telescope Institute, 3700 San Martin Drive,
	 Baltimore, MD 21218, USA\\
	 $^{5}$ National Radio Astronomy Observatory,520 Edgemont Road,
	 Charlottesville, VA22903-2475,USA}

\begin{document}

\maketitle

%AAAAAAAAAAAAAAAAAAAAAAAAAAAAAAAAAAAAAAAAAAAAAAAAAAAAAAAAAAAAAAAAAAA
\begin{abstract}
In a recently completed survey of the stellar populations
properties of LINERS and LINER/HII Transition Objects (TOs), we have
identified a numerous class of galactic nuclei which stand out because
of their conspicuous $10^{8-9}$ yr populations, traced by high order
Balmer absorption lines and other stellar indices. These objects were
called ``Young-TOs'', since they all have TO-like emission line
ratios. In this paper we extend this previous work, which concentrated
on the nuclear properties, by investigating the radial variations of
spectral properties in Low Luminosity Active Galactic Nuclei (LLAGN).
Our analysis is based on high signal to noise long-slit spectra in the
3500--5500 \AA\ interval for a sample of 47 galaxies. The data probe
distances of typically up to 850 pc from the nucleus with a resolution
of $\sim 100$ pc ($\sim 1^{\prime\prime}$) and $S/N \sim 30$. Stellar
population gradients are mapped by the radial profiles of absorption
line equivalent widths and continuum colours along the slit. These
variations are further analyzed by means of a decomposition of each
spectrum in terms of template galaxies representative of very young
($\le 10^7$ yr), intermediate age ($10^{8-9}$ yr) and old ($10^{10}$
yr) stellar populations.

This study reveals that Young-TOs also differ from Old-TOs and
Old-LINERs in terms of the spatial distributions of their stellar
populations and dust. Specifically, our main findings are: (1)
Significant stellar population gradients are found almost exclusively
in Young-TOs. (2) The intermediate age population of Young-TOs,
although heavily concentrated in the nucleus, reaches distances of up
to a few hundred pc from the nucleus. Nevertheless, the Half Width at
Half Maximum of its brightness profile is more typically 100 pc or
less. (3) Objects with predominantly old stellar populations present
spatially homogeneous spectra, be they LINERs or TOs. (4) Young-TOs
have much more dust in their central regions than other LLAGN. (5) The
B-band luminosities of the central $\la 1$ Gyr population in Young-TOs
are within an order of magnitude of $M_B = -15$, implying masses of
order $\sim 10^7$--$10^8$ M$_\odot$. This population was 10--100 times
more luminous in its formation epoch, at which time young massive
stars would have completely outshone any active nucleus, unless the
AGN too was brighter in the past.
\end{abstract}

\begin{keywords} galaxies: active - galaxies: Seyfert - galaxies: 
stellar content - galaxies: nuclei - galaxies: statistics
\end{keywords}
%AAAAAAAAAAAAAAAAAAAAAAAAAAAAAAAAAAAAAAAAAAAAAAAAAAAAAAAAAAAAAAAAAAA

%%%SEC%%%SEC%%%SEC%%%SEC%%%SEC%%%SEC%%%SEC%%%SEC%%%SEC%%%SEC%%%SEC%%%
\section{Introduction}

\label{sec:Introduction}

Low luminosity active galactic nuclei (LLAGN) are the most common form
of activity in the nearby universe. Their proximity allows us to
sample their properties on linear scales which are not accessible for
farther AGN populations like Seyferts and quasars. At optical--UV
wavelengths, however, this advantage is compensated by the difficulty
in isolating the light from these intrinsically weak nuclei out of a
dominant stellar ``background''.

In a series of papers, we have been working our own contribution to
this field.  In Paper I (Cid Fernandes \etal 2004a) we analyzed ground
based nuclear optical spectra of a sample of 51 LINERS and TOs, while
Paper II (Gonz\'alez Delgado \etal 2004) complements this data set
with archive HST/STIS spectra of 28 nearby LLAGN. These samples cover
nearly half of the LLAGN in the survey of Ho, Filippenko \& Sargent
1997 (hereafter HFS97). Our focus throughout this series is on the
stellar populations of LLAGN, with the ultimate goal of establishing
the role of stellar processes in the physics of these objects.

The main result of Papers I and II is that we have uncovered a very
strong relation between the nuclear stellar population and the gas
excitation, as measured by [OI]/H$\alpha$, the most important
diagnostic line ratio in LLAGN. The relation is in the sense that
virtually all systems containing strong populations of $\sim 1$ Gyr or
less have [OI]/H$\alpha \le 0.25$, while nuclei dominated by older
stars span the full range in [OI]/H$\alpha$ (up to nearly 1).  In
other words, virtually all systems with relatively young stellar
populations have TO-like emission line spectra, whereas older nuclei
can have either TO or LINER-like line ratios. This finding lead us to
introduce a combined stellar population and emission line
classification into 4 types: Young-TOs, Old-TOs,
Old-LINERs and Young-LINERs. This latter class is extremely
rare.

This relation between line-ratios and stellar population is analogous
to the one found in Seyfert 2s, where nuclei with strong circumnuclear
starbursts tend to have relatively small values of [OIII]/H$\beta$ and
HeII/H$\beta$, while systems dominated by old stars can reach larger
values of these line ratios (Cid Fernandes \etal 2001 and references
therein). It is thus tempting to interpret Young-TOs as low-luminosity
analogs of starburst + Seyfert 2 composites, where the relatively low
excitation is explained by the starburst contribution to H$\beta$,
which dilutes [OIII]/H$\beta$ and HeII/H$\beta$. Intriguingly,
however, Young-TOs are substantially older than the starbursts around
Seyfert 2 nuclei, many of which are just a few Myr old, as deduced by
the detection of O and WR stars. While Papers I and II revealed a
surprisingly large number of systems containing $10^8$--$10^9$ yr
populations, massive young stars of the type often found in Seyfert 2s
seem to be rare in LLAGN. The analogy between Young-TOs and starburst
+ Seyfert 2 composites thus rests upon the hypothesis of the existence
of a population of massive stars which remains essentially undetected
at optical wavelengths. Clearly, further work is necessary to clarify
the precise nature of the connection between stellar and gaseous
properties in LLAGN.

One type of study which has been carried out for Seyferts is the
mapping of stellar populations based on spatially resolved
spectroscopy. Variations of absorption line equivalent widths
($W_\lambda$) and colours ($C_\lambda$) as a function of distance from
the nucleus were mapped by means of long slit spectroscopy by Cid
Fernandes, Storchi-Bergmann \& Schmitt (1998), Boisson \etal (2000),
Gon\'zalez Delgado, Heckman \& Leitherer (2001), Joguet \etal (2001).
These variations can be transformed into stellar population profiles,
as in the study by Raimann \etal (2003), who found that star-formation
in starburst + Seyfert 2s composites, although concentrated in the
central regions, is not confined to the nucleus, but spread over the
inner $\sim 1$ kpc. Spatial gradients in spectral indices are
also useful to detect the presence of a central continuum source,
which dilutes the nuclear $W_\lambda$'s with respect to off-nuclear
positions. Both a compact nuclear starburst and an AGN featureless
continuum can produce this effect, but in Seyfert 2s the papers above
showed that significant dilution only occurs when a starburst is
present in the innermost extraction.

Spatially resolved spectroscopy of LLAGN has so far been limited to
relatively few studies (eg, Cid Fernandes \etal 1998). While these
previous works advanced our comprehension of individual sources, the
small number of objects, differences in spectral coverage, data
quality and method of analysis prevents us from drawing general
conclusions about the radial distribution of stellar populations in
LINERs and TOs.  In this third paper we take advantage of our recently
completed spectroscopic survey to extend this type of study to a large
sample of LLAGN. Variations of spectral properties with distance from
the nucleus are mapped with the general goal of investigating the
relation between spatial gradients, emission line and nuclear stellar
population properties. In particular, we aim at evaluating the spatial
distribution of intermediate age populations, a distinguishing feature
of Young-TOs.

In \S\ref{sec:Data} we describe the data set and present examples of
our spatially resolved spectra. In
\S\ref{sec:GradentsInStellarProperties} we investigate the spatial
variations of a set of spectral properties and quantify these
gradients by means of suitable empirical indices.  These gradients are
further analyzed in \S\ref{sec:Analysis} with the goal of producing
estimates of the sizes, luminosities, masses and extinction of the
intermediate stellar population in the central regions of Young-TOs.
These estimates provide useful hints on the past and future history of
these sources.  Finally, \S\ref{sec:Conclusions} summarizes our
results.
%%%SEC%%%SEC%%%SEC%%%SEC%%%SEC%%%SEC%%%SEC%%%SEC%%%SEC%%%SEC%%%SEC%%%

%%%SEC%%%SEC%%%SEC%%%SEC%%%SEC%%%SEC%%%SEC%%%SEC%%%SEC%%%SEC%%%SEC%%%
\section{Data}

\label{sec:Data}

The data employed in this paper have been described in Paper
I. Briefly, we have collected long-slit spectra in the 3500--5500 \AA\
range for 60 galaxies selected out of the HFS97 survey.  Observations
were carried out at the 2.5 m Nordic Optical Telescope with a
$1^{\prime\prime}$ slit-width and the Kitt Peak National Observatory
2.1 m telescope with a $2^{\prime\prime}$ slit. Our survey differs
from that of our mother sample in two main aspects: wavelength
coverage and spatial resolution. The information encoded in the region
bluewards of 4200 \AA, not covered by HFS97, has been explored in
previous papers in this series. Here we concentrate on the analysis of
the spatial information in this data set.

\subsection{Extractions}

In order to map spectral gradients, spectra were extracted in several
positions along the slit.  Extractions for the KPNO spectra were made
at every $2.34^{\prime\prime}$ (3 pixels) out to at least $\theta= \pm
4.7^{\prime\prime}$, but the seeing was 2--3$^{\prime\prime}$ (FWHM).
For the NOT spectra, which constitute 83\% of the data analysed here,
we have used $1.13^{\prime\prime}$ (6 pixels) long extractions out to
at least $\theta= 4.5^{\prime\prime}$ from the nucleus in both
directions. These narrow extractions approximately match the angular
resolution of our typical NOT observations, which were made under
sub-arcsecond seeing. Outside this central region wider extractions
were used if necessary to ensure enough signal.

The signal-to-noise ratio in each extraction was estimated from the
rms fluctuation in the 4789--4839 \AA\ interval. Galaxies with
$(S/N)_{\lambda4800} \la 15$ at angular distances $\le
4.7^{\prime\prime}$ from the nucleus were deemed to have insufficient
useful spatial coverage and discarded from the analysis. Our cleaned
sample contains 47 objects, including 4 normal galaxies and 1
Starburst nucleus.  In the nuclear extractions $(S/N)_{\lambda4800}$
varies between 31 and 88 with a median of 51. Outside the nucleus, the
median $(S/N)_{\lambda4800}$ decreases from 45 at $\theta = \pm
2.3^{\prime\prime}$ to 31 at $\pm 4.5^{\prime\prime}$. The $S/N$ in
the 4010--4060 \AA\ interval is typically $0.5
(S/N)_{\lambda4800}$. All 521 extractions were dereddened by Galactic
extinction using the Cardelli, Clayton \& Mathis (1989) law and the
$A_B$ values of Schlegel, Finkbeiner \& Davis (1998).  We note
that the Kitt Peak the observations (7 galaxies) were taken under
non-photometric conditions. This, however affects only the absolute
flux scale, not the shape of the spectrum, as we verified comparing
spectra of objects taken both in photometric and non-photometric
nights. The single result reported in this paper which is affected by
this problem is the luminosity of the central young population in NGC
404 (\S\ref{sec:SizesAndLuminosities_TemplateDecomposition}), which is
likely underestimated.

The distances to the LLAGN in this sample vary between $d = 2.4$ and
70.6 Mpc, with a median of 24.1 Mpc. At these distances, $\theta=
4.5^{\prime\prime}$ corresponds to projected radii $r = 52$--1540 pc,
with a median of 526 pc, while our nuclear extractions correspond to
11--204 pc in radius (median $= 85$ pc). The spatial regions sampled
by these observations are therefore smaller than the ones in our
studies of Seyfert 2s (eg, Raimann \etal 2003), which sampled the
inner few kpc with a resolution of $\sim 300$ pc.

\subsection{Sample properties}

\label{sec:SampleProperties}

Table \ref{tab:sample} lists our sample, along with the useful spatial
coverage in both angular ($\theta_{\rm out}$) and linear ($r_{\rm
out}$) units, nuclear and off-nuclear $S/N$, linear scale, position
angle and a summary of spectral properties.

The emission line classification from HFS97 is listed in column 8 of
Table \ref{tab:sample}. As in Papers I and II, we prefer to classify
LLAGN as either strong or weak-[OI] emitters (column 11), with a
dividing line at [OI]/H$\alpha = 0.25$. These two classes differ only
slightly from the LINER and TO classes of HFS97, and better represent
the combined distributions of emission line and stellar population
properties of LLAGN.  Throughout this paper LINERs and TOs are used as
synonyms of strong and weak-[OI] sources respectively.

Paper I introduced a stellar population characterization scheme
defined in terms of four classes: $\eta = Y$, $I$, $I/O$ and $O$
(column 9). The $Y$ class denotes objects with a dominant young
starburst. The only object in our sample which fits this class is the
WR-galaxy NGC 3367, which is not a LLAGN but is kept in the analysis
for comparison purposes.  Nuclei with strong intermediate age
($10^8$--$10^9$ yr) populations, easily identified by High Order
Balmer absorption Lines (HOBLs; H8$\lambda3889$ and higher) and
diluted metal lines, are classed as $\eta = I$, while nuclei dominated
by old stars are attributed a $\eta = O$ class, and $\eta = I/O$
denotes intermediate cases. Not surprisingly, it is sometimes hard to
decide where to fit a galaxy in this classification scheme. The best
example of this sort of problem is NGC 772, which contains both young,
intermediate age and old components (Paper I). Despite the weak HOBLs
in its spectrum, we chose to tag it as $\eta = I$.

A simpler (but still useful) classification scheme is to group
$\eta = Y$ and $I$ objects as ``Young'' and $\eta = I/O$ and $O$
objects as ``Old''. As an objective criterion for this classification
we use the value of the equivalent width of the CaII K line in the
nucleus: $W_K^{\rm nuc} \le 15$ \AA\ for Young systems and larger for
Old ones (column 10).  The use of this equivalent width as an
indicator of the evolutionary status of the stellar population is
justified because the AGN contribution to these continuum is these
sources negligible (Papers I and II).  These two classes are paired
with the [OI]/H$\alpha$ class to produce our combined stellar
population and emission line classification into Young/Old-TO/LINER,
listed in the last column of Table \ref{tab:sample}.

Of the 42 LLAGN in our sample, 13 fit our definition of strong-[OI]
sources and 29 are weak-[OI] sources, while the stellar populations
types are split into 16 Young and 28 Old systems. The combined
emission line and stellar population statistics are: 14 Young-TOs, 2
Young-LINERs, 11 Old-LINERs and 15 Old-TOs. Note that Young-LINER is a
practically non-existent category, as the overwhelming majority of
Young systems are weak-[OI] emitters.

It is worth pointing out that Young-TOs in this sample are
on-average closer than other LLAGN. The distances to Young-TOs span
the $d = 2.4$--35.6 Mpc range, with a median of 16.8 Mpc, while for
other LLAGN $14.3 \le d \le 70.6$ Mpc, with a median of 31.6 Mpc. This
tendency is already present in Paper I and in the HFS97 survey, from
which we culled our sample. In principle one expects that radial
variations of spectral properties due to the presence of a compact
central source will be harder to detect for more distant objects, due
to the increasing contribution of bulge light to the nuclear
extraction.  This potential difficulty, coupled with the trend
discussed above may lead to a bias in the sense that radial gradients
would be easier to detect in Young-TOs because of their smaller
distances. We do not believe this effect has a strong impact on the
conclusions of this paper, given that there is still a substantial
overlap in distances of Young-TOs and other LLAGN.  This issue is
further discussed in \S\ref{sec:radial_profiles} and
\S\ref{sec:dilution}.

%%%TAB%%%TAB%%%TAB%%%TAB%%%TAB%%%TAB%%%TAB%%%TAB%%%TAB%%%TAB%%%TAB%%%
\begin{table*}
\begin{centering}
\begin{tabular}{lrrrccrrrrrr}
\multicolumn{12}{c}{Observations and Sample Properties}\\ \hline
Galaxy               &
$\theta_{\rm out}$ [$^{\prime\prime}$] &
$r_{\rm out}$ [pc]   &
pc/$^{\prime\prime}$ &
$(S/N)_{\rm nuc}$    &
$(S/N)_{\rm out}$    &
P.A. [$^\circ$]      &
Type                 &
$\eta$               &
$W_K^{\rm nuc}$      &
[OI]                 &
class         \\ 
(1)                  &
(2)                  &
(3)                  &
(4)                  &
(5)                  &
(6)                  &
(7)                  &
(8)                  &
(9)                  &
(10)                 &
(11)                 &
(12)          \\ \hline
%# Galaxy  ang_out  rad_out pc/arcsec S/N_nuc S/N_out       PA   HFS_class       eta   **WKnuc** W/S[OI] Cid-Class
NGC 0266  &    4.5  &  1364   &   303   &  40   &  19   & 100.4  &  L1.9       &  O    & 18.8 &  S  &  Old-LINER \cr
NGC 0315  &    5.6  &  1799   &   319   &  52   &  24   &  94.3  &  L1.9       &  O    & 17.0 &  S  &  Old-LINER \cr
NGC 0404$^\star$  &   11.7  &   136   &    12   &  53   &  24    &   0    &  L2         &  I    &  9.8 &  W  &  Young-TO \cr
NGC 0410  &    4.5  &  1544   &   342   &  61   &  24   &  90.5  &  T2:        &  O    & 17.6 &  W  &  Old-TO \cr
NGC 0521  &    4.5  &  1465   &   325   &  51   &  14   & 124.3  &  T2/H:      &  O    & 17.9 &  W  &  Old-TO \cr
NGC 0718  &    4.5  &   468   &   104   &  44   &  26   & 137.8  &  L2         &  I    & 13.1 &  W  &  Young-TO \cr
NGC 0772  &    6.8  &  1070   &   158   &  57   &  19   & 127.1  &  H/T2:      &  I    & 11.6 &  W  &  Young-TO \cr
NGC 0841  &    4.5  &  1301   &   288   &  65   &  17   &  51.8  &  L1.9:      &  I    & 14.9 &  S  &  Young-LINER \cr
NGC 1052  &    4.5  &   389   &    86   &  68   &  41   & 167.1  &  L1.9       &  O    & 17.3 &  S  &  Old-LINER \cr
NGC 1161  &    6.8  &   850   &   126   &  56   &  21   &  37.6  &  T1.9:      &  O    & 19.0 &  W  &  Old-TO \cr
NGC 2681  &    7.3  &   473   &    64   &  41   &  28   &  81.5  &  L1.9       &  I    & 12.3 &  W  &  Young-TO \cr
NGC 2685  &    7.9  &   620   &    79   &  40   &  29   & 167.3  &  S2/T2:     &  I/O  & 18.7 &  W  &  Old-TO \cr
NGC 3166  &   10.7  &  1143   &   107   &  45   &  20   & 188.4  &  L2         &  I/O  & 15.9 &  S  &  Old-LINER \cr
NGC 3245  &    4.5  &   485   &   108   &  66   &  33   & 264.5  &  T2:        &  I/O  & 15.2 &  W  &  Old-TO \cr
NGC 3627  &   13.4  &   427   &    32   &  50   &  32   & 222.6  &  T2/S2      &  I    & 11.6 &  W  &  Young-TO \cr
NGC 3705  &    5.6  &   465   &    82   &  31   &  18   & 218.2  &  T2         &  I    & 14.8 &  W  &  Young-TO \cr
NGC 4150  &    4.5  &   212   &    47   &  42   &  23   & 267.6  &  T2         &  I    & 12.6 &  W  &  Young-TO \cr
NGC 4438  &    7.3  &   597   &    81   &  40   &  25   & 224.4  &  L1.9       &  I/O  & 17.8 &  S  &  Old-LINER \cr
NGC 4569  &    4.5  &   367   &    81   &  57   &  27   & 228.7  &  T2         &  I    &  5.0 &  W  &  Young-TO \cr
NGC 4736  &   17.5  &   364   &    21   &  50   &  34   & 324.1  &  L2         &  I    & 12.9 &  W  &  Young-TO \cr
NGC 4826  &    4.5  &    90   &    20   &  42   &  39   & 250.5  &  T2         &  I    & 14.4 &  W  &  Young-TO \cr
NGC 5005  &   11.6  &  1194   &   103   &  53   &  18   & 286.6  &  L1.9       &  I    & 14.6 &  S  &  Young-LINER \cr
NGC 5377  &    6.8  &  1017   &   150   &  48   &  18   & 288.6  &  L2         &  I    &  8.7 &  W  &  Young-TO \cr
NGC 5678  &    7.3  &  1265   &   173   &  53   &  30   & 149.5  &  T2         &  I    &  8.7 &  W  &  Young-TO \cr
NGC 5921  &    4.5  &   551   &   122   &  46   &  13   & 188.3  &  T2         &  I    & 11.2 &  W  &  Young-TO \cr
NGC 5970  &    7.3  &  1123   &   153   &  32   &  20   & 234.1  &  L2/T2:     &  I/O  & 18.4 &  W  &  Old-TO \cr
NGC 5982  &    6.2  &  1163   &   188   &  56   &  40   & 311.5  &  L2::       &  O    & 18.1 &  S  &  Old-LINER \cr
NGC 5985  &    4.5  &   857   &   190   &  38   &  11   & 308.8  &  L2         &  I/O  & 18.9 &  S  &  Old-LINER \cr
NGC 6340$^\star$  &    9.4  &   998   &   107   &  61   &  18   &   0    &  L2         &  O    & 20.0 &  S  &  Old-LINER \cr
NGC 6384  &    4.5  &   582   &   129   &  40   &  16   & 211.3  &  T2         &  I/O  & 18.6 &  W  &  Old-TO \cr
NGC 6482  &    7.3  &  1859   &   254   &  74   &  28   & 115.8  &  T2/S2::    &  O    & 18.8 &  W  &  Old-TO \cr
NGC 6500  &    4.5  &   868   &   192   &  50   &  14   & 197.2  &  L2         &  I/O  & 15.8 &  W  &  Old-TO \cr
NGC 6501  &    4.5  &   866   &   192   &  59   &  20   & 240.2  &  L2::       &  O    & 16.8 &  S  &  Old-LINER \cr
NGC 6503  &   13.4  &   395   &    30   &  40   &  26   &  10.1  &  T2/S2:     &  I    &  9.5 &  W  &  Young-TO \cr
NGC 6702  &    4.5  &  1373   &   304   &  55   &  17   & 335.5  &  L2::       &  O    & 18.1 &  S  &  Old-LINER \cr
NGC 6703$^\star$  &    9.4  &  1629   &   174   &  55   &  33   &   0    &  L2::       &  O    & 18.5 &  S  &  Old-LINER \cr
NGC 6951  &    4.5  &   527   &   117   &  39   &  30   &  0     &  S2/L       &  I/O  & 16.4 &  W  &  Old-TO \cr
NGC 7177$^\star$  &   11.7  &  1032   &    88   &  55   &  27   &   0    &  T2         &  I/O  & 16.6 &  W  &  Old-TO \cr
NGC 7217$^\star$  &   11.7  &   908   &    78   &  48   &  25   &   0    &  L2         &  O    & 19.2 &  W  &  Old-TO \cr
NGC 7331$^\star$  &   16.4  &  1136   &    69   &  57   &  39   &   0    &  T2         &  O    & 18.0 &  W  &  Old-TO \cr
NGC 7626  &    4.5  &   997   &   221   &  71   &  35   & 122.6  &  L2::       &  O    & 18.1 &  W  &  Old-TO \cr
NGC 7742$^\star$  &   11.7  &  1259   &   108   &  47   &  32   &   0    &  T2/L2      &  I/O  & 17.1 &  W  &  Old-TO \cr \hline
NGC 3367  &    4.5  &   953   &   211   &  88   &  15   &  203.6 &  H          &  Y    &  2.6 & -  &  -  \cr
NGC 0224  &    7.9  &    27   &     3   &  60   &  66   &   66.5 &  normal     &  O    & 17.5 & -  &  -  \cr
NGC 0628  &    6.2  &   292   &    47   &  40   &  25   &  167.3 &  normal     &  I/O  & 16.1 & -  &  -  \cr
NGC 1023  &    7.9  &   402   &    51   &  59   &  35   &  293.7 &  normal     &  O    & 19.4 & -  &  -  \cr
NGC 2950  &    5.6  &   637   &   113   &  44   &  24   &   48.5 &  normal     &  O    & 17.4 & -  &  -  \cr \hline
%# Output of Table1_Paper3_LLAGN.LaTeX_format.for
%# Cid@lynx - 01/Jan/2004
% WK added by hand - Cid@Lagoa - 31/Aug/2004
\end{tabular}
\end{centering}
\caption{Col.\ (1): Galaxy name; Cols.\ (2) and (3): Useful angular
and linear coverage. Col.\ (4): Angular scale. Cols.\ (5) and (6): S/N
at 4800 \AA\ at nucleus an outer extractions. Col.\ (7). Slit position
angle. Col.\ (8): Spectral type according to HFS97. Col.\ (9) Stellar
population category (Paper I). Col.\ (10): Equivalent width of
the CaII K band at the nucleus, in \AA.  Col.\ (11): W = Weak-[OI]
(ie, [OI]/H$\alpha \le 0.25$), S = Strong-[OI] ([OI]/H$\alpha >
0.25$). Col.\ (12): Combined emission line and stellar population
class.  Objects marked with a $\star$ were observed at KPNO.}
\label{tab:sample}
\end{table*}
%%%TAB%%%TAB%%%TAB%%%TAB%%%TAB%%%TAB%%%TAB%%%TAB%%%TAB%%%TAB%%%TAB%%%

\subsection{Spatially resolved spectra: Examples and first impressions}

\label{sec:SpatiallyResolvedSpectra}

%%%FIG%%%FIG%%%FIG%%%FIG%%%FIG%%%FIG%%%FIG%%%FIG%%%FIG%%%FIG%%%FIG%%%
\begin{figure*}
\psfig{file=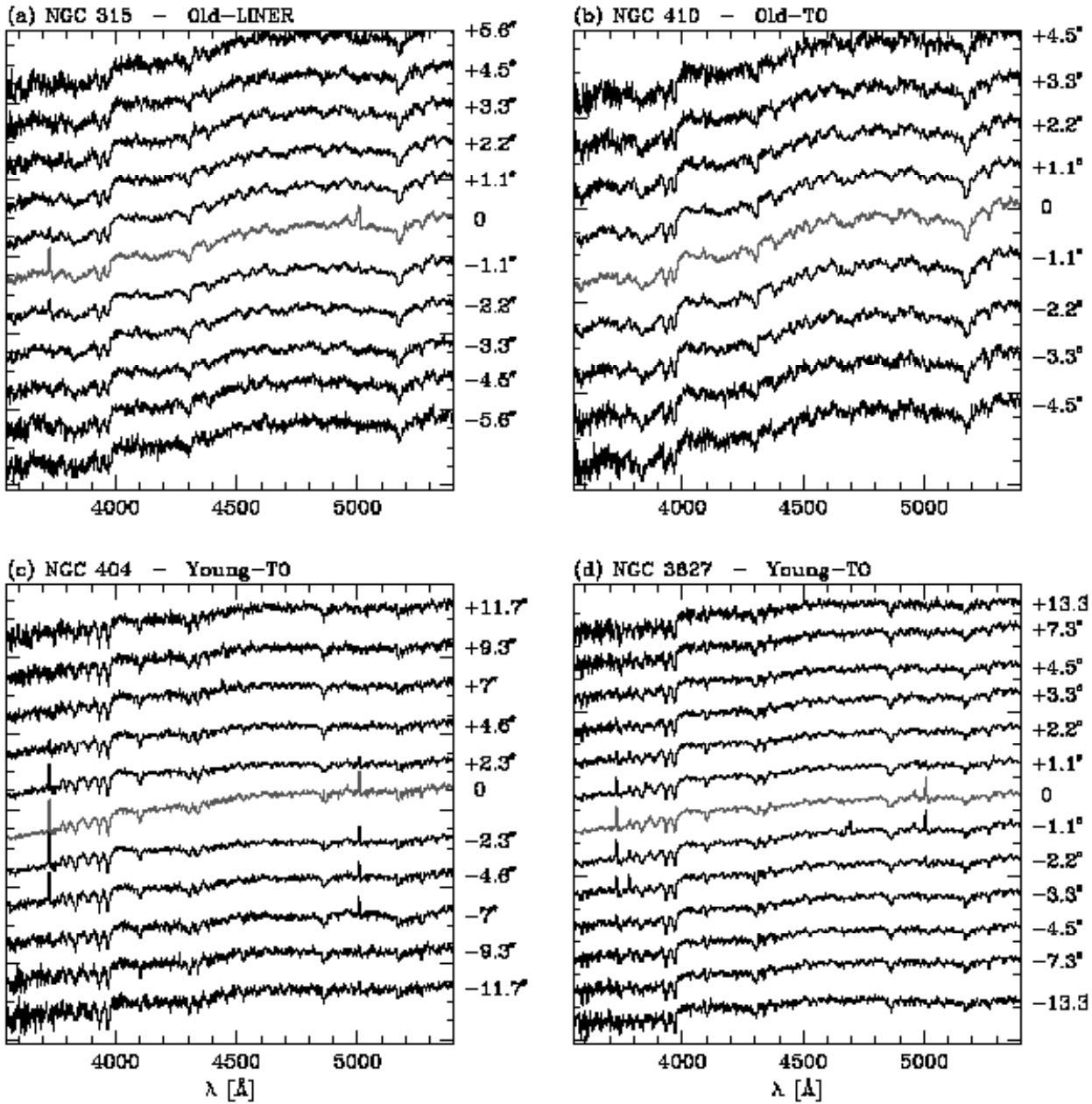,width=\textwidth,bbllx=0,bblly=200,bburx=480,bbury=710}
%\psfig{file=Fig_offnuc_spec_examples1.eps2,width=13cm,bbllx=140,bblly=160,bburx=520,bbury=710}
%\resizebox{\textwidth}{!}{\includegraphics{Fig_offnuc_spec_examples1.eps2}}
\caption{Examples of spatially resolved spectra of LLAGN. Spectra have
been normalized and shifted for clarity. The nuclear spectrum is drawn
with a thicker line. Labels in the right indicate the angular distance
from the nucleus.}
\label{fig:offnuc_spectra_examples1}
\end{figure*}

\begin{figure*}
\psfig{file=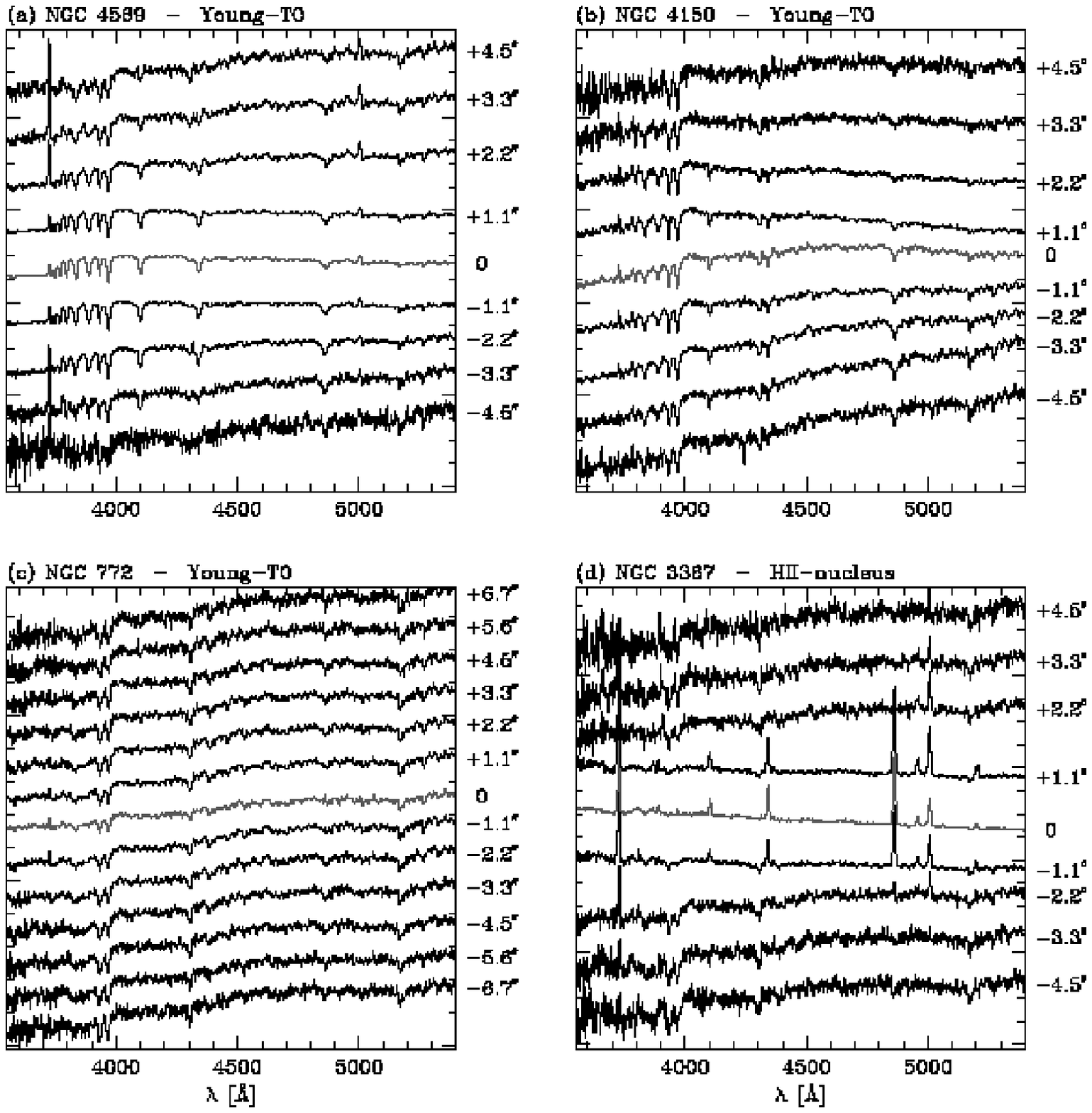,width=\textwidth,bbllx=0,bblly=200,bburx=480,bbury=710}
%\psfig{file=Fig_offnuc_spec_examples2.eps2,width=13cm,bbllx=140,bblly=160,bburx=520,bbury=710}
%\resizebox{\textwidth}{!}{\includegraphics{Fig_offnuc_spec_examples2.eps2}}
\caption{As Fig.\ \ref{fig:offnuc_spectra_examples1}.}
\label{fig:offnuc_spectra_examples2}
\end{figure*}                                           
%%%FIG%%%FIG%%%FIG%%%FIG%%%FIG%%%FIG%%%FIG%%%FIG%%%FIG%%%FIG%%%FIG%%%

Figures \ref{fig:offnuc_spectra_examples1} and
\ref{fig:offnuc_spectra_examples2} illustrate spatially resolved
spectra for a representative subset of the galaxies in our
sample. Spatial gradients in spectral properties will be analyzed in
detail in the remainder of this paper, but some results are evident
from a simple visual inspection of these figures.

\begin{enumerate}

\item First, in objects like the Old-LINER NGC 315 the off-nuclear
spectra look virtually identical to the nuclear spectrum, implying a
high spatial uniformity of the stellar populations. The only
noticeable gradient is in the emission lines, which are concentrated
in the nucleus.

\item Second, the strongest gradients are found in systems with
conspicuous HOBLs (eg, NGC 4150, NGC 4569). As noted above, these are
nearly all weak-[OI] sources. This combination of youngish stellar
population and [OI]/H$\alpha \le 0.25$ fits our definition of
Young-TOs.

\item Third, although HOBLs, when present, are stronger in the central
extraction, they are {\it not} confined to the nucleus. This is
clearly seen in the cases of NGC 4150 and NGC 4569, where HOBLs still
show up in extractions more than 3\arcsec\ away from the nucleus.
Given that the seeing in these observations was typically better than
1\arcsec, we conclude that the ``HOBLs region'' is {\it spatially
extended}. 

\item As is typical of LLAGN, emission lines are generally weak. In
fact, many objects show no sign of important diagnostic lines like
H$\beta$ and [OIII]$\lambda$5007 even in the nucleus.  The measurement
of emission lines requires careful subtraction of the starlight, which
we postpone to a future communication.

\end{enumerate}

%%%SEC%%%SEC%%%SEC%%%SEC%%%SEC%%%SEC%%%SEC%%%SEC%%%SEC%%%SEC%%%SEC%%%

%%%SEC%%%SEC%%%SEC%%%SEC%%%SEC%%%SEC%%%SEC%%%SEC%%%SEC%%%SEC%%%SEC%%%
\section{Stellar Population Gradients}

\label{sec:GradentsInStellarProperties}

A convenient way to map spatio-spectral variations is to compute
profiles of absorption features and continuum colours along the slit
(eg, Cid Fernandes \etal 1998; Raimann \etal 2003).  From Papers I and
II we know that an AGN continuum contributes very little (if anything)
to our ground based optical spectra. Any significant variation
detected in these properties can thus be confidently attributed to
variations in the stellar populations.

In Paper I we have measured an extensive set of stellar population
indices in different systems.  In this paper we will use the
equivalent widths of the CaII K line ($W_K$), the G band ($W_G$), MgI
($W_{Mg}$) and $W_C$ (a ``pseudo equivalent width'' centered in the
continuum just to the blue of H9) plus the $C_{3660} \equiv 3660/4020$
and $C_{5313} \equiv 5313/4020$ continuum colours, all measured in
Bica's system.  The 4000 \AA\ break index of Balogh \etal (1999),
$D_n(4000)$, is also used, but only for illustrative purposes. The
$W_C$ index works as a direct tracer of HOBLs: Spectra with clearly
visible HOBLs all have $W_C < 3.5$ \AA, while spectra dominated by old
populations ($\sim 10^{10}$ yr) have larger $W_C$ due to a blend of
metal lines.  As shown in Paper I, $W_K$, which is a much stronger and
thus more robust feature, is also a good (albeit indirect) tracer of
the intermediate age populations responsible for the HOBLs.

All these indices are highly correlated (Paper I). Their radial
behaviors, however, need not be the same. For instance, a compact blue
source such as young or intermediate age starburst should produce a
larger dilution at the nucleus of the bluer indices, like $W_K$, than
of the redder ones, such as $W_{Mg}$. The comparison of the $W_K$ and
$W_{Mg}$ profiles may thus allow inferences about the nuclear stellar
population.

We have measured these indices automatically for all 521 extractions
analyzed in this work following the recipes outlined in Paper I for
Bica's indices and Balogh \etal (1999) for $D_n(4000)$.  This served
as a further test of the objective pseudo continuum definition
proposed in Paper I. After visual inspection of the results, we have
judged that only in 18 spectra (3\% of the total) the pseudo continuum
deserved corrections.  Uncertainties in all spectral indices were
estimated by means of Monte Carlo simulations. The typical
uncertainties at $\theta= \pm 4.5^{\prime\prime}$, the outermost
extractions in many of our sources, are 0.5--1 \AA\ for all equivalent
widths, and 0.04 for $C_{3660}$, $C_{5313}$ and $D_n(4000)$. Indices
for the nuclear extractions are 2--3 times more accurate.

\subsection{Radial Profiles of Stellar Indices}

\label{sec:radial_profiles}

%%%FIG%%%FIG%%%FIG%%%FIG%%%FIG%%%FIG%%%FIG%%%FIG%%%FIG%%%FIG%%%FIG%%%
\begin{figure}
\psfig{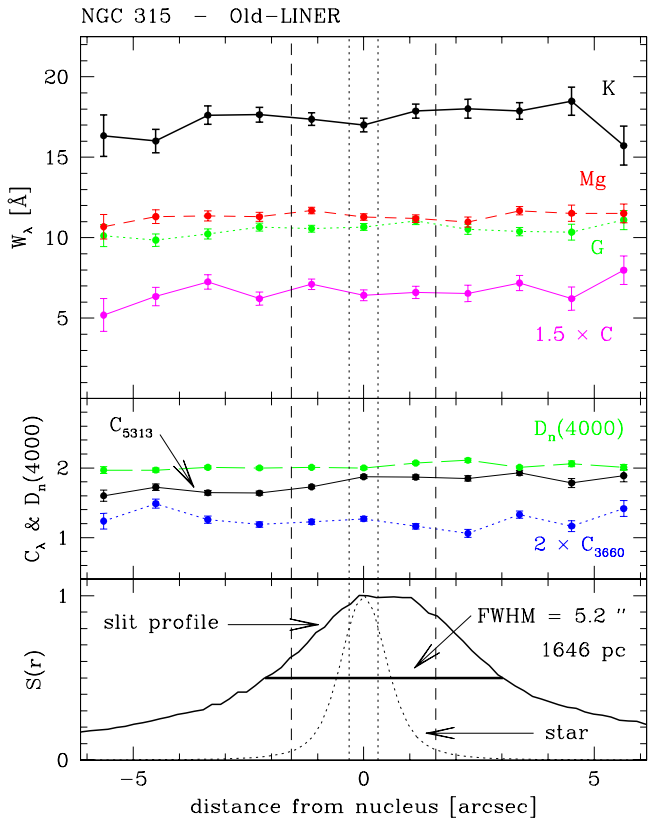}
\caption{Spatial variations of stellar population indices for NGC
  315. {\it Top:} Radial profiles of $W_K$ (black, thick solid line),
  $W_C$ (magenta, thin line), $W_G$ (green, dotted line) and $W_{Mg}$
  (red, dashed line). Note that $W_C$ has been multiplied by 1.5 for
  clarity.  {\it Middle:} Radial profile of the 3660/4020 (blue,
  dotted line) and 5313/4020 (black, solid line) colours, and
  $D_n(4000)$ (green, dashed line). The 3660/4020 colour is multiplied
  by 2 in the plot. {\it Bottom:} Surface brightness at 4200 \AA\ (in
  flux units) along the slit, normalized to $S(r=0) = 1$. The FWHM of
  the slit profile is marked as a thick line segment, and listed at
  the top right in both arcsec and pc. The dotted line shows the
  instrumental profile, corresponding to a star observed in the same
  night. Dotted and dashed vertical lines indicate projected distances
  of $\pm 100$ and 500 pc from the nucleus respectively.}
\label{fig:grad_NGC0315}
\end{figure}                                           
%%%FIG%%%FIG%%%FIG%%%FIG%%%FIG%%%FIG%%%FIG%%%FIG%%%FIG%%%FIG%%%FIG%%%

\begin{figure}
\psfig{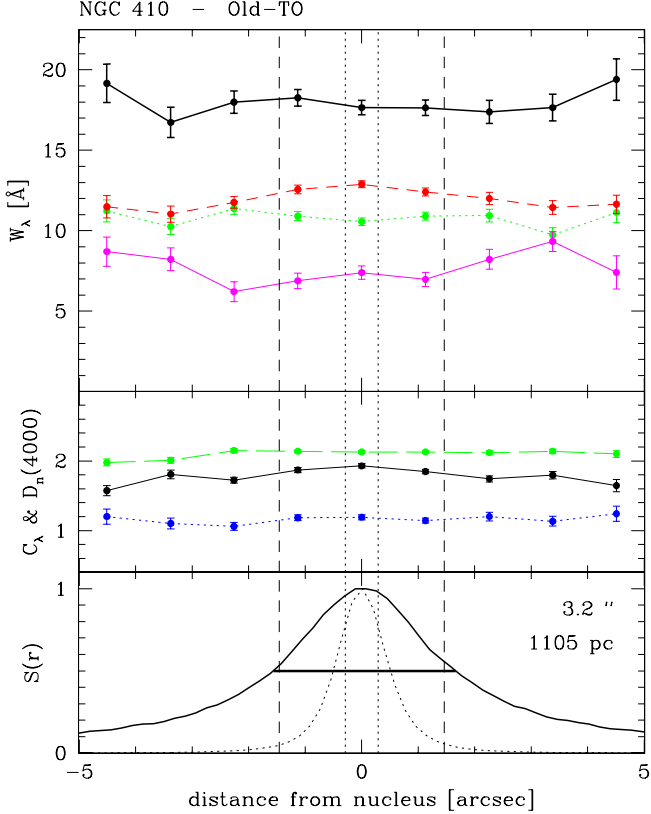}
\caption{As Figure \ref{fig:grad_NGC0315}, but for NGC 410.}
\label{fig:grad_NGC0410}
\end{figure}

\begin{figure}
\psfig{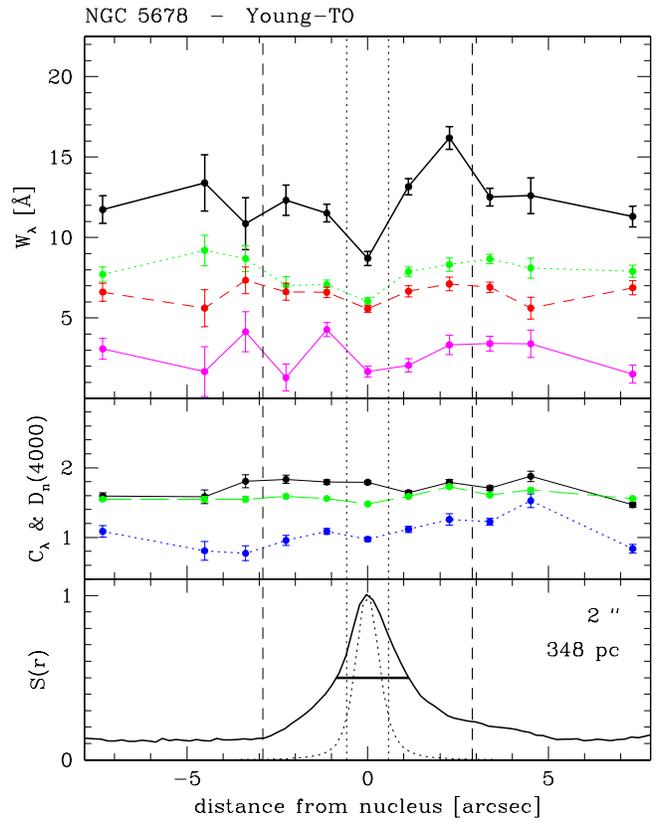}
\caption{As Figure \ref{fig:grad_NGC0315}, but for NGC 5678.}
\label{fig:grad_NGC5678}
\end{figure}

\begin{figure}
\psfig{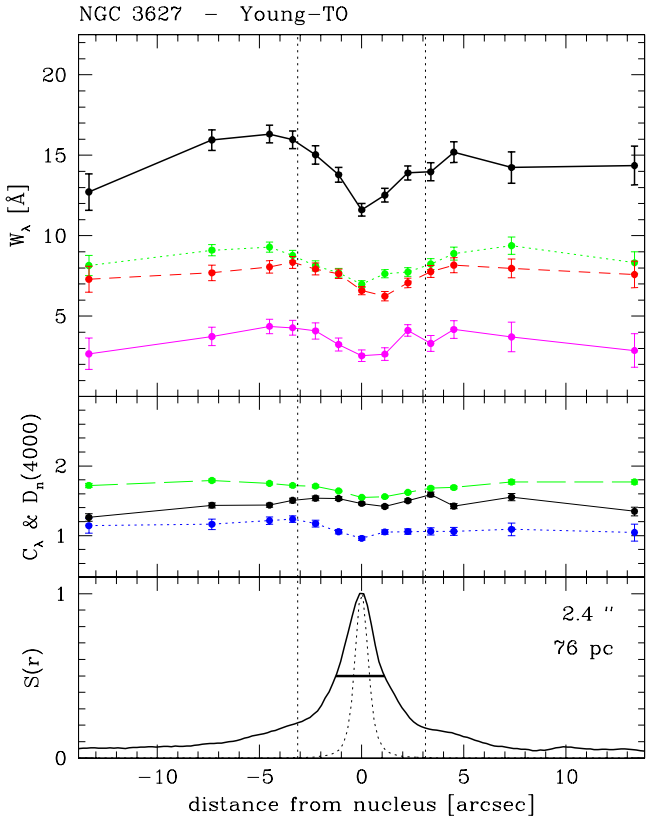}
\caption{As Figure \ref{fig:grad_NGC0315}, but for NGC 3627.}
\label{fig:grad_NGC3627}
\end{figure}

\begin{figure}
\psfig{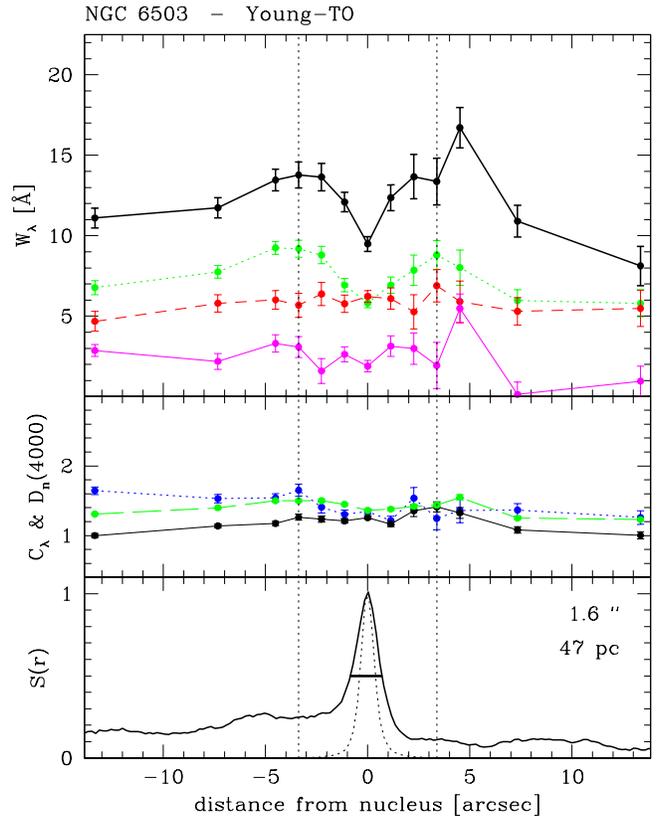}
\caption{As Figure \ref{fig:grad_NGC0315}, but for NGC 6503.}
\label{fig:grad_NGC6503}
\end{figure}

\begin{figure}
\psfig{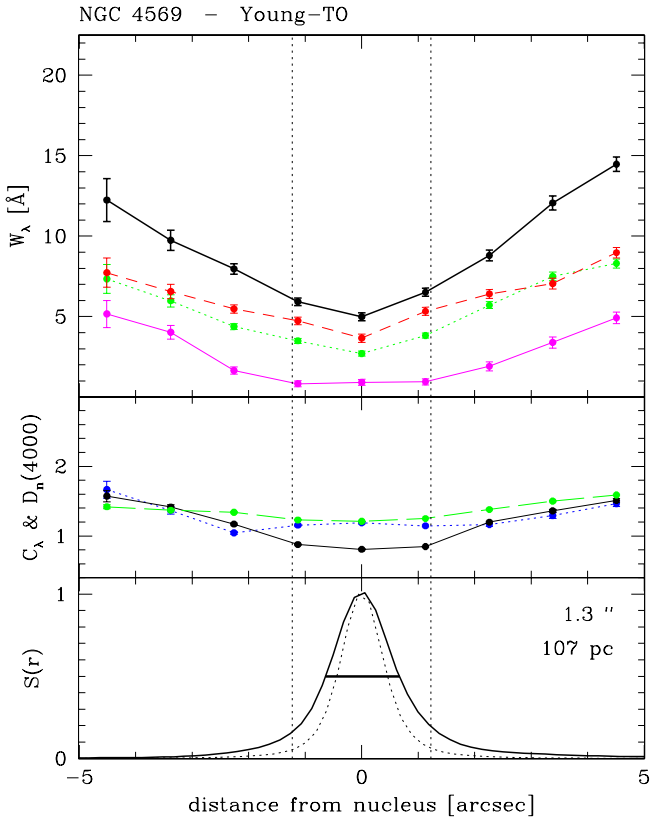}
\caption{As Figure \ref{fig:grad_NGC0315}, but for NGC 4569.}
\label{fig:grad_NGC4569}
\end{figure}

Figures \ref{fig:grad_NGC0315}--\ref{fig:grad_NGC3367} show the
variations of our seven stellar population indices with angular
distance from the nucleus for some illustrative cases.  The top panels
show $W_K$ (black, solid line), $W_C$ (magenta, thin line), $W_G$
(green, dotted line) and $W_{Mg}$ (red, dashed line). The middle
panels show the $C_{3660}$ (blue, dotted line) and $C_{5313}$ (black,
solid line) colours, plus the $D_n(4000)$ profile (green, dashed
line). The slit brightness profile $S(r)$ at $\lambda = 4200$ \AA\ is
plotted in the bottom panel to give an idea of the light
concentration.  The thick line segment marks the FWHM of $S(r)$; its
value is listed in the top right in both angular and linear units. A
stellar profile is also plotted to illustrate the spatial
resolution. Vertical dotted and dashed lines indicate projected
distances of $\pm 100$ and $\pm 500$ pc from the nucleus respectively.

The examples in figures \ref{fig:grad_NGC0315}--\ref{fig:grad_NGC3367}
were chosen to illustrate the variety of radial profiles found in the
sample. In a first cut, the $W_\lambda$ profiles may be grouped in
three categories: 

\begin{enumerate}

\item Flat (eg., NGC 305 and NGC 410), 

\item centrally peaked (eg., NGC 7742),

\item ``diluted'' profiles (eg., NGC 3627, NGC 4569).

\end{enumerate}

Most objects studied here have either flat or diluted $W_\lambda$
profiles.  In NGC 6951 and NGC 7742, the peaked appearance of
$W_\lambda(r)$ is due to circum-nuclear star-forming rings which
appear in our outermost extractions (P\'erez \etal 2000).  Outside
these rings, the absorption lines rise up again, like in NGC 1097 and
other ringed galaxies studied by Cid Fernandes \etal (1998).

The main focus of our analysis throughout the rest of this paper will
be on nature and properties of the source of dilution in LLAGN with
diluted profiles.  These profiles cannot be explained in terms of
metallicity gradients, as this should produce peaked profiles.  The
drop in $W_\lambda$ towards the nucleus in these galaxies is thus
clearly the result of dilution of the metallic features by a centrally
concentrated stellar population which is younger than that a few
arcseconds away from the nucleus. The most dramatic example of this
effect is seen in the starburst galaxy NGC 3367, where the young
starburst appears only in the three central extractions (figure
\ref{fig:grad_NGC3367}). We note in passing that, at $d = 43.6$ Mpc,
this galaxy is one of the most distant in our sample, well above the
median distance of 27.9 Mpc. Yet, its $W_\lambda$ gradients are
clearly mapped with our data, which shows that the worries raised in
\S\ref{sec:SampleProperties} about possible distance related biases
and not justified in practice.  Similar comments apply to NGC 5678
(figure \ref{fig:grad_NGC5678}, $d = 35.6$ Mpc) and NGC 772 (figure
\ref{fig:grad_NGC0772}, $d = 32.6$ Mpc).

In LLAGN with diluted profiles, the diluting agent could in principle
also be a young starburst, but, as shown in Papers I and II, in only
$\sim 10\%$ of LLAGN such a young component contributes with more than
10\% of the flux at 4020 \AA\ in our ground-based nuclear spectra.
For most objects, the radial dilution is caused mainly by an
intermediate age population, which appears far more frequently and in
much larger strengths. These populations are easily recognized by
their weak metal lines and deep HOBLs, as seen, for instance, in NGC
3627 and NGC 4569 (figures \ref{fig:offnuc_spectra_examples1},
\ref{fig:offnuc_spectra_examples2}, \ref{fig:grad_NGC3627} and
\ref{fig:grad_NGC4569}).

%%%FIG%%%FIG%%%FIG%%%FIG%%%FIG%%%FIG%%%FIG%%%FIG%%%FIG%%%FIG%%%FIG%%%

\begin{figure}
\psfig{file=Fig_grads_n7742.eps2,width=9.6cm,bbllx=40,bblly=160,bburx=520,bbury=710}
\caption{As Figure \ref{fig:grad_NGC0315}, but for NGC 7742.}
\label{fig:grad_NGC7742}
\end{figure}                                           

\begin{figure}
\psfig{file=Fig_grads_n4150.eps2,width=9.6cm,bbllx=40,bblly=160,bburx=520,bbury=710}
\caption{As Figure \ref{fig:grad_NGC0315}, but for NGC 4150.}
\label{fig:grad_NGC4150}
\end{figure}

\begin{figure}
\psfig{file=Fig_grads_n772.eps2,width=9.6cm,bbllx=40,bblly=160,bburx=520,bbury=710}
\caption{As Figure \ref{fig:grad_NGC0315}, but for NGC 772.}
\label{fig:grad_NGC0772}
\end{figure}                                           

\begin{figure}
\psfig{file=Fig_grads_n3367.eps2,width=9.6cm,bbllx=40,bblly=160,bburx=520,bbury=710}
\caption{As Figure \ref{fig:grad_NGC0315}, but for NGC 3367, a
starburst galaxy in our comparison sample.}
\label{fig:grad_NGC3367}
\end{figure}                                           
%%%FIG%%%FIG%%%FIG%%%FIG%%%FIG%%%FIG%%%FIG%%%FIG%%%FIG%%%FIG%%%FIG%%%

\begin{figure*}
\psfig{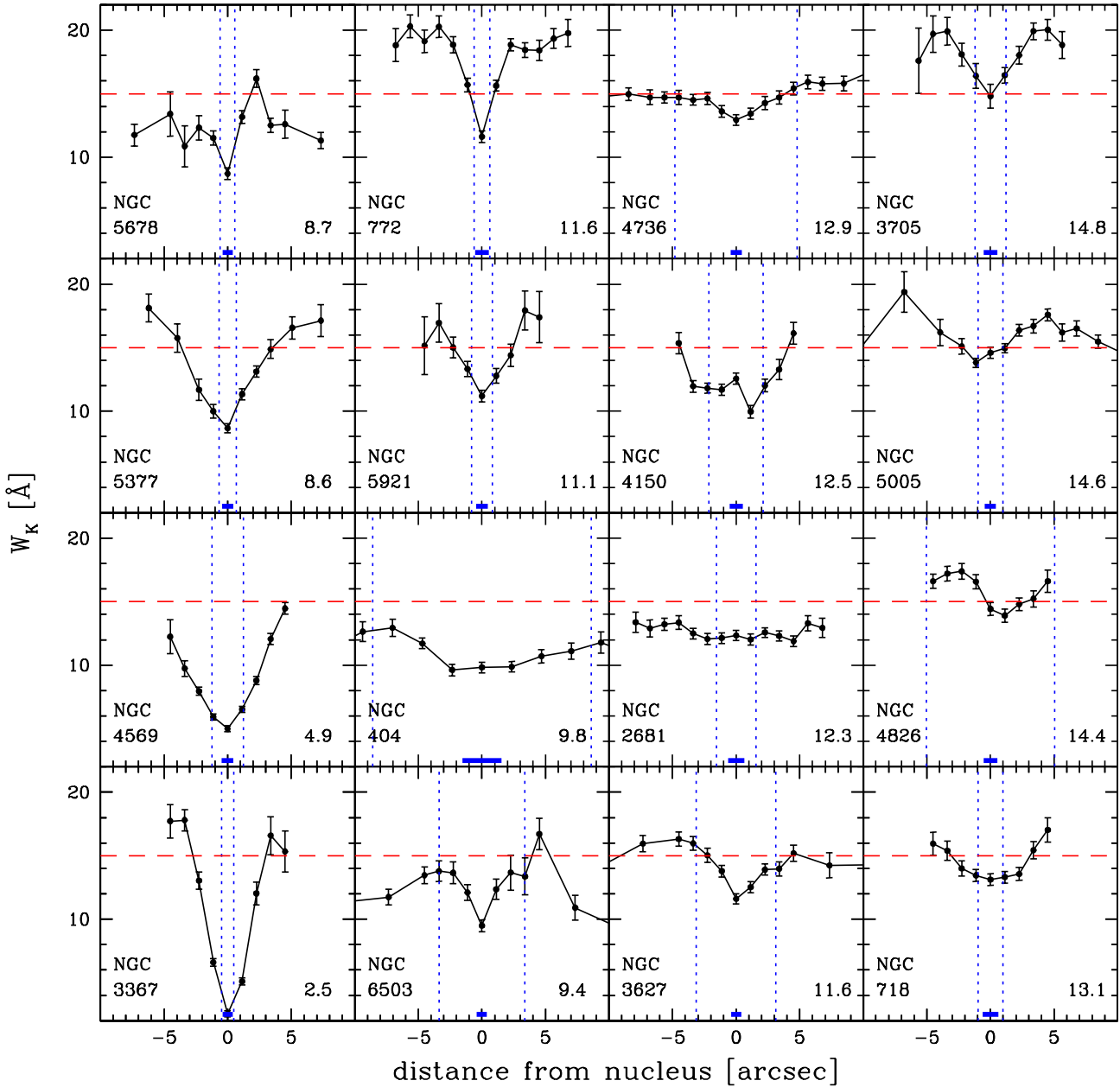}
%\resizebox{\textwidth}{!}{\includegraphics{Fig_WK_profiles1.eps2}}
\caption{Radial profiles of the equivalent width of the Ca II K line
for all galaxies in the sample. Dotted vertical lines mark distances
of $\pm 100$ pc from the nucleus. A horizontal dashed line is drawn at
$W_K = 15$ \AA\ for reference. The thick line segment in the
bottom of each panel indicates the seeing, measured from the FWHM of
star observed in the same night. Objects are sorted by the value of
$W_K$ at the nucleus ($W_K^{\rm nuc}$), indicated in the bottom right
corner of each panel. Galaxies in this figure have $W_K^{\rm nuc}$
between 2.5 (bottom left panel) and 14.8 \AA\ (top right).}
\label{fig:WK_profiles1}
\end{figure*}

Figures \ref{fig:WK_profiles1}--\ref{fig:WK_profiles3} show the $W_K$
profiles for all 47 galaxies in our sample, {\it sorted in an
increasing sequence of nuclear $W_K$ values}. This ordering bears an
excellent correspondence with the profile shapes: Of the first 19
galaxies (from NGC 3367 to NGC 6500), at least 16 have diluted $W_K$
profiles. The exceptions are NGC 2681, NGC 841 and possibly NGC
6500. From NGC 3166 onwards, ie, for $W^{\rm nuc}_K > 15$--16 \AA,
profiles are either centrally peaked, or, more commonly, approximately
flat. This obvious link is examined in quantitative terms in the next
section.

%%%FIG%%%FIG%%%FIG%%%FIG%%%FIG%%%FIG%%%FIG%%%FIG%%%FIG%%%FIG%%%FIG%%%

\begin{figure*}
\psfig{file=Fig_WK_profiles2.eps2,width=13cm,bbllx=100,bblly=160,bburx=500,bbury=710}
%\resizebox{\textwidth}{!}{\includegraphics{Fig_WK_profiles2.eps2}}
\caption{As Figure \ref{fig:WK_profiles1}, but for galaxies with
$W_K^{\rm nuc}$ between 14.9 and 17.8 \AA.}
\label{fig:WK_profiles2}
\end{figure*}                                           

\begin{figure*}
\psfig{file=Fig_WK_profiles3.eps2,width=13cm,bbllx=100,bblly=160,bburx=500,bbury=710}
%\resizebox{\textwidth}{!}{\includegraphics{Fig_WK_profiles3.eps2}}
\caption{As Figure \ref{fig:WK_profiles1}, but for galaxies with
$W_K^{\rm nuc}$ between 18 and 20 \AA.}
\label{fig:WK_profiles3}
\end{figure*}                                           
%%%FIG%%%FIG%%%FIG%%%FIG%%%FIG%%%FIG%%%FIG%%%FIG%%%FIG%%%FIG%%%FIG%%%

\subsection{Gradients in Equivalent Widths}

\label{sec:dilution}

In order to quantify the spatial gradients seen in figures
\ref{fig:grad_NGC0315}--\ref{fig:WK_profiles3} we define a radial
dilution index

\begin{equation}
\label{eq:dilution}
\delta_\lambda = 
       \frac{W^{\rm off}_\lambda - W^{\rm nuc}_\lambda}{W^{\rm off}_\lambda}.
\end{equation}

\noindent which compares nuclear and mean off-nuclear equivalent
widths.  Flat $W_\lambda$ profiles should yield $\delta_\lambda \sim
0$, while $\delta_\lambda < 0$ correspond to centrally peaked profiles
and $\delta_\lambda > 0$ to diluted profiles. Furthermore, if the
nuclear spectrum differs from that in off-nuclear extractions only by
an extra continuum source (or, more precisely, a source with
negligible $W_\lambda$), then $\delta_\lambda$ measures the fractional
contribution of this source to the continuum at $\lambda$ (Cid
Fernandes \etal 1998).

$W^{\rm off}_\lambda$ is defined as the average of $W_\lambda(\theta)$
for extractions centered at $|\theta|$ between 2.2 and
$4.7^{\prime\prime}$ from the nucleus. Note that for the NOT
observations this definition excludes extractions adjacent to the
nucleus, which in some cases are contaminated by nuclear light due to
seeing.  The averaging is carried out weighting by the error in
$W_\lambda$. The uncertainties in the dilution index were evaluated
from standard error propagation. Typical one sigma uncertainties in
$\delta_\lambda$ are 0.1 for $W_C$ and 0.04 for $W_K$, $W_G$ and
$W_{Mg}$.  We have also explored an alternative definition of $W^{\rm
off}_\lambda$ in terms of extractions between $r = 250$ and 750 pc
from the nucleus, but this turned out to yield similar results,
which further demonstrates that our conclusions are not significantly
affected by potential distance-related biases
(\S\ref{sec:SampleProperties}).

Table \ref{tab:dilution} lists the resulting values of
$\delta_\lambda$. Gradients are considered to be significant whenever
$|\delta_K| > 10\%$, which corresponds to a $\sim 2.5$ sigma detection
limit. According to this criterion, significantly diluted profiles
($\delta_K > 10\%$) occur in 13 of the 42 LLAGN in our sample, while
only 3 have significantly peaked profiles ($\delta_K < -10\%$).
Spatially homogeneous stellar populations therefore prevail among
LLAGN, accounting for $\sim 60\%$ of our sample.

\subsubsection{Relations between $W_\lambda$-gradients, emission line and
nuclear stellar population properties}

In figure \ref{fig:dil_X_EW} we investigate the relation between
dilution and nuclear stellar population by plotting $\delta_\lambda$
against $W^{\rm nuc}_\lambda$ for $W_C$, $W_K$, $W_G$ and $W_{Mg}$.
The vertical dotted lines in this plot are the same ones used in Paper
I to approximately distinguish objects with significant intermediate
age populations (those with $W_C \la 3.5$, $W_K \la 15$, $W_G \la 9$
and $W_{Mg} \la 9$ \AA, which are classed as $\eta = I$) from those
dominated by older populations ($\eta = I/O$ and $O$).  Figure
\ref{fig:dil_X_EW} shows that these dividing lines also segregate
objects with significant dilution from those without.  Focusing on the
$W^{\rm nuc}_K = 15$ \AA\ limit, which separates Young from Old
sources in our simple classification scheme, we find that 12 out of
the 13 objects with $\delta_K > 10\%$ fall into the Young category,
the exception being NGC 3245, which, with $W_K^{\rm nuc} = 15.2 \pm
0.3$ \AA, sits right at the border line between Young and Old sources.
In other words, {\it galaxies with significant radial gradients in
their stellar populations contain intermediate age populations in
their nuclei.} The converse is also true, as at least 12 out of 16
Young-LLAGN have diluted profiles. This is the same result found in
figures \ref{fig:WK_profiles1}--\ref{fig:WK_profiles3}, where we see
that virtually every galaxy with $W_K^{\rm nuc} \la 15$ \AA\ has a
diluted $W_K$ profile.

%%%TAB%%%TAB%%%TAB%%%TAB%%%TAB%%%TAB%%%TAB%%%TAB%%%TAB%%%TAB%%%TAB%%%
\begin{table*}
\begin{centering}
\begin{tabular}{lrrrrrrr}
\multicolumn{8}{c}{Radial Dilution, Colour Gradients and Sizes}\\ \hline
NGC                       &
$\delta_C$ [\%]           &
$\delta_K$ [\%]           &
$\delta_G$ [\%]           &
$\delta_{Mg}$ [\%]        &
$\delta_V$ [mag]          &
$R_S$ [$^{\prime\prime}$] &
$R_S$ [pc]                \\ 
(1)  &
(2)  &
(3)  &
(4)  &
(5)  &
(6)  &
(7)  &
(8)  \\ \hline
%# NGC    da(C) err    da(K)  err    da(G)  err   da(Mg)  err	  dAV(C5313)   err	HWHM in " and pc
0266 &    2$\pm$ 9 &    6$\pm$ 3 &    3$\pm$ 3 &   -6$\pm$ 3 &	  0.18 $\pm$ 0.07 &	 1.4 & 435 \cr
0315 &    4$\pm$ 6 &    3$\pm$ 3 &   -3$\pm$ 2 &    1$\pm$ 2 &	  0.17 $\pm$ 0.04 &	 2.6 & 822 \cr
0404 &    7$\pm$15 &    6$\pm$ 4 &   15$\pm$ 3 &    9$\pm$ 6 &	  0.54 $\pm$ 0.04 &	 2.1 &  25 \cr
0410 &    8$\pm$ 6 &    1$\pm$ 3 &    2$\pm$ 3 &  -11$\pm$ 3 &	  0.29 $\pm$ 0.05 &	 1.6 & 552 \cr
0521 &    0$\pm$ 9 &    3$\pm$ 4 &    2$\pm$ 3 &    1$\pm$ 3 &	  0.09 $\pm$ 0.07 &	 1.1 & 371 \cr
0628 &  -11$\pm$13 &   -4$\pm$ 4 &    2$\pm$ 4 &    0$\pm$ 7 &	  0.22 $\pm$ 0.07 &	 7.2 & 339 \cr
0718 &   21$\pm$ 9 &   12$\pm$ 3 &    5$\pm$ 3 &    0$\pm$ 5 &	  0.24 $\pm$ 0.05 &	 1.4 & 143 \cr
0772 &   39$\pm$ 6 &   39$\pm$ 3 &   21$\pm$ 3 &   13$\pm$ 4 &	 -0.36 $\pm$ 0.05 &	 0.7 & 111 \cr
0841 &   10$\pm$10 &   -3$\pm$ 4 &    0$\pm$ 4 &   -7$\pm$ 5 &	  0.37 $\pm$ 0.06 &	 1.1 & 310 \cr
1023 &  -11$\pm$ 7 &   -2$\pm$ 3 &   -3$\pm$ 3 &   -8$\pm$ 2 &	  0.29 $\pm$ 0.05 &	 1.9 &  97 \cr
1052 &  -30$\pm$12 &   -6$\pm$ 4 &    1$\pm$ 3 &   -8$\pm$ 3 &	  0.79 $\pm$ 0.06 &	 1.3 & 112 \cr
1161 &  -62$\pm$11 &  -17$\pm$ 3 &  -17$\pm$ 3 &  -10$\pm$ 2 &	  0.56 $\pm$ 0.05 &	 1.2 & 151 \cr
2681 &   17$\pm$11 &    1$\pm$ 4 &    3$\pm$ 4 &   -4$\pm$ 6 &	  0.11 $\pm$ 0.04 &	 1.6 & 101 \cr
2685 &  -18$\pm$ 9 &   -4$\pm$ 4 &   -7$\pm$ 3 &  -20$\pm$ 6 &	  0.43 $\pm$ 0.07 &	 2.4 & 188 \cr
2950 &   -8$\pm$ 9 &    2$\pm$ 4 &   -3$\pm$ 3 &  -14$\pm$ 4 &	  0.26 $\pm$ 0.06 &	 1.6 & 184 \cr
3166 &    6$\pm$ 9 &    0$\pm$ 4 &   -6$\pm$ 4 &  -11$\pm$ 5 &	  0.03 $\pm$ 0.05 &	 1.7 & 179 \cr
3245 &   10$\pm$ 5 &   15$\pm$ 2 &   11$\pm$ 2 &   -6$\pm$ 2 &	  0.28 $\pm$ 0.04 &	 1.0 & 110 \cr
3367 &   82$\pm$ 6 &   83$\pm$ 2 &   86$\pm$ 2 &   82$\pm$ 4 &	 -1.86 $\pm$ 0.06 &	 0.4 &  95 \cr
3627 &   38$\pm$ 9 &   23$\pm$ 3 &   18$\pm$ 3 &   16$\pm$ 4 &	 -0.06 $\pm$ 0.04 &	 1.2 &  38 \cr
3705 &   28$\pm$13 &   23$\pm$ 5 &   18$\pm$ 5 &   17$\pm$ 6 &	  0.25 $\pm$ 0.09 &	 0.7 &  59 \cr
4150 &    8$\pm$12 &    3$\pm$ 4 &   11$\pm$ 3 &   20$\pm$ 7 &	 -0.73 $\pm$ 0.05 &	 2.0 &  93 \cr
4438 &  -17$\pm$10 &  -11$\pm$ 4 &   -7$\pm$ 3 &   -9$\pm$ 3 &	  1.14 $\pm$ 0.05 &	 2.8 & 227 \cr
4569 &   70$\pm$ 6 &   52$\pm$ 2 &   57$\pm$ 3 &   47$\pm$ 4 &	 -1.24 $\pm$ 0.03 &	 0.6 &  53 \cr
4736 &   11$\pm$ 9 &   12$\pm$ 3 &   10$\pm$ 3 &    5$\pm$ 4 &	 -0.10 $\pm$ 0.05 &	 1.1 &  24 \cr
4826 &   23$\pm$ 9 &   11$\pm$ 3 &    8$\pm$ 3 &    2$\pm$ 4 &	 -0.06 $\pm$ 0.05 &	 1.0 &  20 \cr
5005 &   24$\pm$ 8 &   11$\pm$ 3 &   15$\pm$ 3 &   19$\pm$ 3 &	  0.17 $\pm$ 0.05 &	 1.1 & 109 \cr
5377 &   38$\pm$ 7 &   37$\pm$ 3 &   20$\pm$ 3 &   -3$\pm$ 5 &	 -0.41 $\pm$ 0.05 &	 0.8 & 116 \cr
5678 &   44$\pm$13 &   34$\pm$ 4 &   27$\pm$ 3 &   17$\pm$ 4 &	  0.03 $\pm$ 0.05 &	 1.0 & 170 \cr
5921 &   34$\pm$ 9 &   29$\pm$ 4 &   22$\pm$ 4 &   16$\pm$ 6 &	 -0.38 $\pm$ 0.07 &	 1.0 & 116 \cr
5970 &    4$\pm$ 8 &   -7$\pm$ 4 &   -2$\pm$ 4 &  -15$\pm$ 7 &	  0.30 $\pm$ 0.07 &	 3.2 & 496 \cr
5982 &   -5$\pm$ 7 &   -4$\pm$ 3 &   -3$\pm$ 3 &   -8$\pm$ 3 &	  0.13 $\pm$ 0.05 &	 1.2 & 227 \cr
5985 &   -4$\pm$12 &    7$\pm$ 4 &    0$\pm$ 4 &   -2$\pm$ 6 &	  0.49 $\pm$ 0.08 &	 1.5 & 294 \cr
6340 &   -9$\pm$ 8 &   -5$\pm$ 3 &   -8$\pm$ 3 &   -9$\pm$ 3 &	  0.96 $\pm$ 0.05 &	 3.1 & 326 \cr
6384 &  -21$\pm$10 &    3$\pm$ 4 &    8$\pm$ 3 &   -9$\pm$ 5 &	  0.49 $\pm$ 0.07 &	 2.4 & 316 \cr
6482 &   -8$\pm$ 6 &   -3$\pm$ 3 &   -6$\pm$ 2 &  -13$\pm$ 2 &	  0.43 $\pm$ 0.04 &	 1.2 & 298 \cr
6500 &   13$\pm$ 7 &    6$\pm$ 3 &   10$\pm$ 3 &   -4$\pm$ 3 &	  0.07 $\pm$ 0.06 &	 1.0 & 199 \cr
6501 &  -47$\pm$13 &    1$\pm$ 3 &    6$\pm$ 2 &   -4$\pm$ 2 &	  0.34 $\pm$ 0.05 &	 1.1 & 220 \cr
6503 &   39$\pm$13 &   32$\pm$ 4 &   34$\pm$ 4 &   -3$\pm$ 8 &	 -0.02 $\pm$ 0.06 &	 0.8 &  23 \cr
6702 &  -22$\pm$10 &   -9$\pm$ 4 &  -16$\pm$ 4 &  -18$\pm$ 4 &	  0.07 $\pm$ 0.06 &	 1.1 & 332 \cr
6703 &   -8$\pm$ 8 &   -2$\pm$ 3 &   -1$\pm$ 3 &   -8$\pm$ 3 &	  0.35 $\pm$ 0.06 &	 2.1 & 369 \cr
6951 & -222$\pm$46 &  -90$\pm$ 8 &  -54$\pm$ 5 &  -50$\pm$ 4 &	  1.21 $\pm$ 0.05 &	 3.7 & 435 \cr
7177 &  -11$\pm$ 9 &   -3$\pm$ 3 &   -2$\pm$ 3 &   -5$\pm$ 4 &	  0.28 $\pm$ 0.05 &	 4.6 & 404 \cr
7217 &   -8$\pm$ 8 &    0$\pm$ 3 &    1$\pm$ 3 &   -9$\pm$ 3 &	  0.53 $\pm$ 0.06 &	 2.9 & 221 \cr
7331 &   -1$\pm$ 7 &    1$\pm$ 3 &    2$\pm$ 3 &   -4$\pm$ 3 &	  0.04 $\pm$ 0.05 &	 3.1 & 215 \cr
7626 &    1$\pm$ 6 &    4$\pm$ 3 &   -2$\pm$ 2 &  -10$\pm$ 2 &	  0.38 $\pm$ 0.05 &	 1.3 & 293 \cr
7742 &  -19$\pm$11 &   -4$\pm$ 4 &   -2$\pm$ 3 &  -10$\pm$ 5 &	  0.59 $\pm$ 0.06 &	 2.2 & 237 \cr \hline
%# Output of Dilution_LLAGN.- Color_Gradient_LLAGN.- & CalcSlitProfileWidth_LaTeX_format.for
%# Cid@Lynx - 28/Jan/2004		
\end{tabular}
\end{centering}
\caption{Cols.\ 2--5: Dilution of the equivalent widths $W_C$, $W_K$,
$W_G$ and $W_{Mg}$. Col.\ 6: Gradient in the $C_{5313}$ colour (see
equation \ref{eq:dAV_C5313}). Cols.\ 7 and 8: HWHM of the flux profile
along the slit, in angular and linear units.}
\label{tab:dilution}
\end{table*}
%%%TAB%%%TAB%%%TAB%%%TAB%%%TAB%%%TAB%%%TAB%%%TAB%%%TAB%%%TAB%%%TAB%%%

%%%FIG%%%FIG%%%FIG%%%FIG%%%FIG%%%FIG%%%FIG%%%FIG%%%FIG%%%FIG%%%FIG%%%
\begin{figure*}
\psfig{file=Fig_dilution_x_Ws.eps2,width=14.2cm,bbllx=70,bblly=150,bburx=520,bbury=710}
%\resizebox{\textwidth}{!}{\includegraphics{Fig_dilution_x_Ws.eps2}}
\caption{Radial gradients in four equivalent widths, measured from the
comparison of nuclear and off-nuclear spectra.  Different symbols
correspond to Young-TOs (filled blue circles), Young-LINERs (open blue
circles), Old-TOs (filled red triangles) and Old-LINERs (open red
triangles). The star indicates the starburst galaxy NGC 3367.  Crosses
in the top right indicate mean error bars. Vertical dotted lines
divide nuclei containing only old stars (large $W_\lambda$) from those
with significant intermediate age populations (small $W_\lambda$).
Note that NGC 6951, which has very negative $\delta_\lambda$'s
due to its star-forming ring, is outside all plot scales.}
\label{fig:dil_X_EW}
\end{figure*}                                           
%%%FIG%%%FIG%%%FIG%%%FIG%%%FIG%%%FIG%%%FIG%%%FIG%%%FIG%%%FIG%%%FIG%%%

Since in Papers I and II we have shown that nearly all nuclei
with weak metal absorption lines are weak-[OI] emitters, we expect
that the strong relation between $\delta_\lambda$ and $W_\lambda^{\rm
nuc}$ seen in figure \ref{fig:dil_X_EW} translates to an equally
strong relation between $\delta_\lambda$ and [OI]/H$\alpha$. This is
confirmed in figure \ref{fig:dil_X_O1Ha}, which shows that all but one
object with $\delta_K > 10\%$ have [OI]/H$\alpha < 0.25$. Two other
weak-[OI] nuclei, NGC 404 and NGC 4150, should probably be included in
the list of sources with diluted profiles. NGC 404 is so close by (2.4
Mpc) that our outer useful extractions do not reach a probable
rise in $W_\lambda$'s for larger radii, if this indeed happens in this
dwarf galaxy.  Dust effects may also be present, as indicated by the
peak in the $C_{5313}$ colour in the nucleus of NGC 404 (figure
\ref{fig:COLOR_profiles1}).  In NGC 4150 the strongest dilution is
seen at $\theta = +1.1^{\prime\prime}$ from the nucleus, and the rise
in $W_\lambda$ seen in our last extractions has a small weight in our
definition of $W^{\rm off}_\lambda$, resulting in a small
$\delta_\lambda$. This asymmetry is associated with the pronounced
nuclear dust lane in this galaxy (Paper II), which is responsible for
its asymmetric $C_{5313}$ profile (figure \ref{fig:grad_NGC4150}).

The only strong-[OI] source with significant radial dilution is NGC
5005 ($\delta_K = 11 \pm 3 \%$, [OI]/H$\alpha = 0.65$). Given that
this nucleus is classified as a L1.9 by HFS97, it is conceivable that
the dilution is caused by a nuclear featureless continuum, as found in
spatially resolved spectroscopy of type 1 Seyferts (Cid Fernandes
\etal 1998). However, none of the other 7 type 1 LLAGN in our sample
exhibits significant dilution. Furthermore, HOBLs are clearly present
in the nuclear spectrum of NGC 5005, so we favor the interpretation
that, as in other objects, dilution is caused mainly by a centrally
concentrated intermediate age population. As noted in Paper II, and
confirmed by our radial dilution analysis, the contribution of a
non-stellar continuum to our ground based spectra is negligible.
Clear signatures of a featureless continuum in LLAGN are only found
under the much higher spatial resolution of HST, and even then they
are rare.

We thus conclude that {\it virtually all sources with radially diluted
metal lines are weak-[OI] emitters}.  Note, however, that the converse
is {\it not} true, as there are several weak-[OI] objects with either
flat or, more rarely, peaked $W_\lambda$ profiles.  These non-diluted
weak-[OI] nuclei are dominated by old stellar populations, as deduced
from their strong metal lines (figure \ref{fig:dil_X_EW}, Paper I).

To summarize, combining the relations between dilution, stellar
population and emission line properties we find that significant
stellar populations gradients are found almost exclusively in
Young-TOs, ie, objects with weak [OI] {\it and} a conspicuous
intermediate age nuclear stellar population. Old-TOs and Old-LINERs,
on the other hand, tend to have spatially uniform stellar
populations. These strong relations can be visualized comparing the
location of different symbols in figures \ref{fig:dil_X_EW} and
\ref{fig:dil_X_O1Ha}.

%%%FIG%%%FIG%%%FIG%%%FIG%%%FIG%%%FIG%%%FIG%%%FIG%%%FIG%%%FIG%%%FIG%%%
\begin{figure}
\psfig{file=Fig_dilution_x_O1Ha.eps2,width=7.5cm,bbllx=30,bblly=120,bburx=500,bbury=710}
%\resizebox{9cm}{!}{\includegraphics{Fig_dilution_x_O1Ha.eps2}}
\caption{Radial dilution in the K line against the [OI]/H$\alpha$
emission line ratio (extracted from HFS97). Symbols as in figure
\ref{fig:dil_X_EW}.  The horizontal line shows the [OI]/H$\alpha =
0.25$ taxonomical frontier which separates strong from weak-[OI]
nuclei.  Objects to the right of the vertical line at $\delta_K =
10\%$ are those with significant dilution.}
\label{fig:dil_X_O1Ha}
\end{figure}                                           
%%%FIG%%%FIG%%%FIG%%%FIG%%%FIG%%%FIG%%%FIG%%%FIG%%%FIG%%%FIG%%%FIG%%%

\subsubsection{$W_\lambda$-gradients and the colour of the nuclear source}

Another result of the $W_\lambda(r)$ analysis is that the spatial
dilution, when significant, tends to be larger for shorter
wavelengths, which implies that the diluting agent is {\it bluer} than
the off-nuclear stellar population. This is illustrated in figure
\ref{fig:dilutions_x_lambda}, where we plot the dilution in the K line
(central $\lambda = 3930$ \AA) against the dilution in $W_C$ ($\lambda
= 3816$ \AA), $W_G$ ($\lambda = 4301$ \AA) and $W_{Mg}$ ($\lambda =
5176$ \AA). For LLAGN with $\delta_K \ga 10\%$ the dilution follows a
wavelength sequence: $\delta_C > \delta_K > \delta_G >
\delta_{Mg}$. (Deviations from this sequence are all within the
uncertainties in $\delta_\lambda$).  Some objects with clear gradients
in K show little, if any, dilution in MgI (eg, NGC 3245 and NGC 6503).
In NGC 772, NGC 4569 and other objects, the colour profiles confirm the
existence of the blue nuclear component inferred from the behavior of
$\delta_\lambda$ for different lines. In others, however,
$C_\lambda(r)$ shows little variation (eg, NGC 3627) or even slightly
redder colours in the nucleus (NGC 3245), {\it contrary} to the
inference from the absorption line gradients. As discussed below, this
apparent contradiction is due to dust in the central regions of these
galaxies.

%%%FIG%%%FIG%%%FIG%%%FIG%%%FIG%%%FIG%%%FIG%%%FIG%%%FIG%%%FIG%%%FIG%%%
\begin{figure}
\psfig{file=Fig_dilution_x_lambda.eps2,width=7.5cm,bbllx=30,bblly=120,bburx=500,bbury=710}
%\resizebox{9cm}{!}{\includegraphics{Fig_dilution_x_lambda.eps2}}
\caption{Dilution in $W_C$ (diamonds), $W_G$ (crosses) and $W_{Mg}$
(filled circles) against the dilution in $W_K$. The diagonal line
simply marks $y = x$.  All sources in this plot are LLAGN. Error bars
have been omitted for clarity.  }
\label{fig:dilutions_x_lambda}
\end{figure}                                           
%%%FIG%%%FIG%%%FIG%%%FIG%%%FIG%%%FIG%%%FIG%%%FIG%%%FIG%%%FIG%%%FIG%%%

\subsection{Colour gradients and extinction}

\label{sec:ColourGradients}

Colour gradients carry information on the variations of stellar
populations and extinction across a galaxy.  Our $C_{3660}$ colour
brackets the region containing the 4000 \AA\ break and Balmer jump,
while $C_{5313}$ is roughly equivalent to $B-V$.  Because of the
larger wavelength interval involved (5313 to 4020 \AA) and the absence
of spectral discontinuities in this range, $C_{5313}$ is the more
reddening sensitive of the two indices. One must nevertheless bear in
mind that a $C_{5313}(r)$ profile cannot be trivially transformed into
an extinction profile without a simultaneous analysis of stellar
population variations.

\begin{figure*}
\psfig{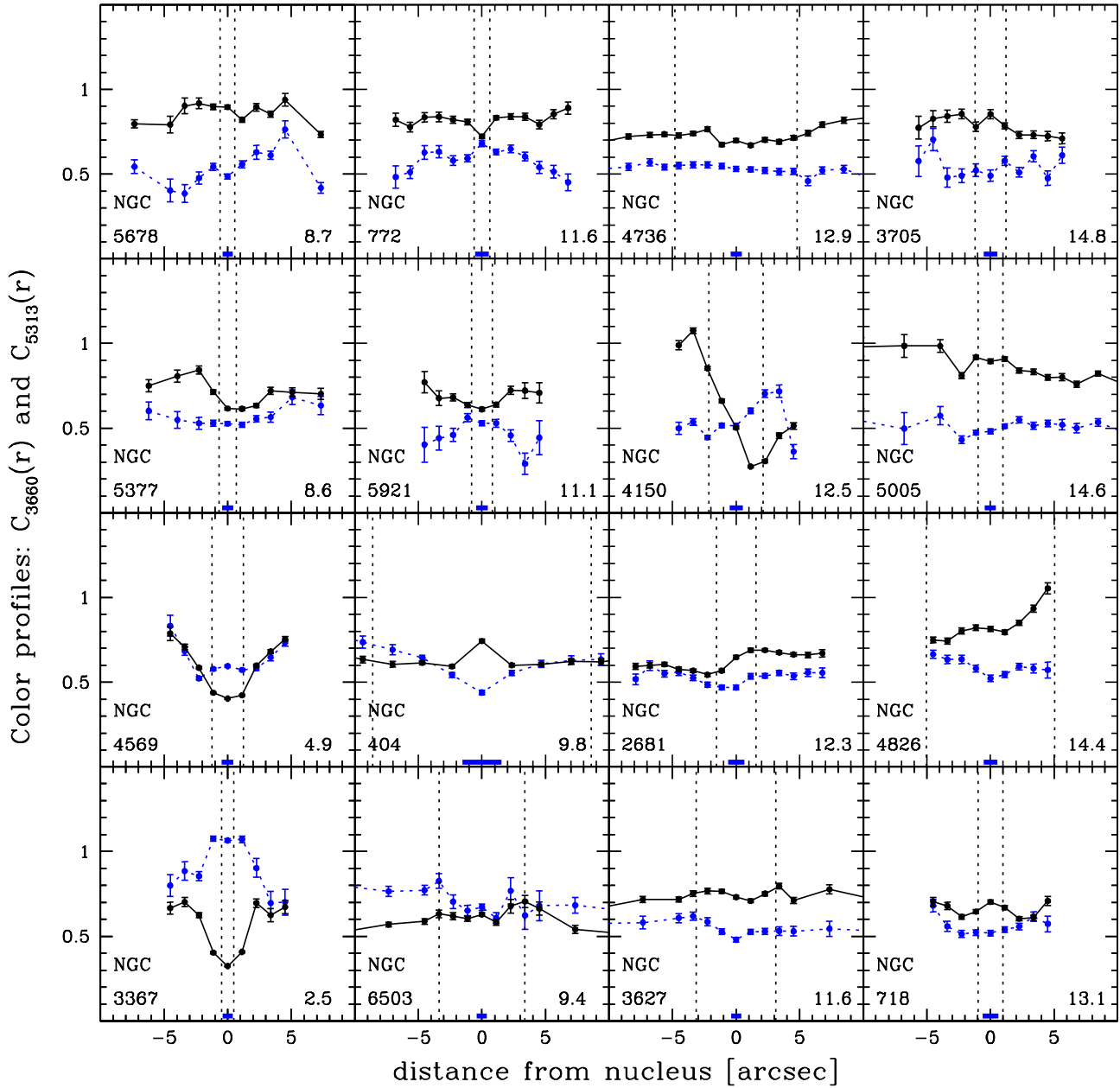}
%\resizebox{\textwidth}{!}{\includegraphics{Fig_COLOR_profiles1.eps2}}
\caption{Radial profiles of two continuum colours: $C_{3660} =
3660/4020$ (dotted, blue line) and 5313/4020 (solid, black line). Note
that the values of $C_{5313}$ have been divided by 2 for plotting
purposes.  Dotted vertical lines mark distances of $\pm 100$ pc from
the nucleus.  Objects are sorted by the value of $W_K$ at the nucleus,
indicated in the bottom right corner of each panel. Galaxies in this
figure have $W_K^{\rm nuc}$ between 2.5 (bottom left panel) and 14.8
\AA\ (top right).}
\label{fig:COLOR_profiles1}
\end{figure*}

Figures \ref{fig:COLOR_profiles1}--\ref{fig:COLOR_profiles3} show the
$C_{3660}$ (dotted, blue line) and $C_{5313}$ (solid, black line)
colour profiles, also ordered according to $W_K^{\rm nuc}$.  Centrally
peaked $C_{5313}(r)$ profiles are apparently rare among galaxies with
$W_K^{\rm nuc} \la 15$ \AA\ (figure \ref{fig:COLOR_profiles1}), with
exceptions (eg., NGC 404, NGC 718 and NGC 3245). This type of profile
appears more often in figures \ref{fig:COLOR_profiles2} and
\ref{fig:COLOR_profiles3}, which contain galaxies with $W_K^{\rm nuc}
\ga 15$ \AA.

%%%FIG%%%FIG%%%FIG%%%FIG%%%FIG%%%FIG%%%FIG%%%FIG%%%FIG%%%FIG%%%FIG%%%

\begin{figure*}
\psfig{file=Fig_COLOR_profiles2.eps2,width=13cm,bbllx=100,bblly=160,bburx=500,bbury=710}
%\resizebox{\textwidth}{!}{\includegraphics{Fig_COLOR_profiles2.eps2}}
\caption{As Figure \ref{fig:COLOR_profiles1}, but for galaxies with
$W_K^{\rm nuc}$ between 14.9 and 17.8 \AA.}
\label{fig:COLOR_profiles2}
\end{figure*}                                           

\begin{figure*}
\psfig{file=Fig_COLOR_profiles3.eps2,width=13cm,bbllx=100,bblly=160,bburx=500,bbury=710}
%\resizebox{\textwidth}{!}{\includegraphics{Fig_COLOR_profiles3.eps2}}
\caption{As Figure \ref{fig:COLOR_profiles1}, but for galaxies with
$W_K^{\rm nuc}$ between 18 and 20 \AA.}
\label{fig:COLOR_profiles3}
\end{figure*}                                           
%%%FIG%%%FIG%%%FIG%%%FIG%%%FIG%%%FIG%%%FIG%%%FIG%%%FIG%%%FIG%%%FIG%%%

In order to examine colour gradients in more quantitative terms we
compute the ratio $C_{5313}^{\rm nuc} / C_{5313}^{\rm off}$ between
the values of $C_{5313}$ in the nucleus and a mean off-nuclear colour,
defined as the weighted average of extractions between $|\theta| =
2.2$ and 4.7$^{\prime\prime}$ (as done for $W_\lambda^{\rm off}$ in
\S\ref{sec:dilution}). This ratio can be transformed into the index

\begin{equation}
\label{eq:dAV_C5313}
\delta_V = 5.98  \log 
  \left( \frac{C_{5313}^{\rm nuc}}{C_{5313}^{\rm off}} \right)
\end{equation}

\ni which measures by how many V-band magnitudes one has to deredden
the nuclear spectrum to make it match the off-nuclear $C_{5313}$
colour. The coefficient in this equation comes from assuming the
Cardelli \etal (1989) extinction curve with $R_V = 3.1$, which we do
throughout this paper.  $\delta_V < 0 $, which indicates a bluening
towards the nucleus, is henceforth referred to as a ``blue gradient'',
while $\delta_V > 0 $ is called a ``red gradient''.  Colour gradients
were also examined by fitting the nuclear spectrum with a combination
of off-nuclear spectra plus reddening, which yields the nuclear
extinction relative to that of the off-nuclear extractions. This
method gives essentially identical results to those based solely on
the $C_{5313}$ colour, with a mean off-set of just 0.04 mag and rms
difference of 0.16 mag between the two $\delta_V$ estimates.

Figure \ref{fig:dAV5313} compares $\delta_V$ with $\delta_K$,
$W_K^{\rm nuc}$ and [OI]/H$\alpha$. The figure confirms that red
gradients are more common for objects with $W_K^{\rm nuc} > 15$ \AA,
ie, among Old-LLAGN, which also tend to have flat $W_\lambda$
profiles.  In fact, the plot shows that {\it all} objects with
$W_K^{\rm nuc} > 15$ \AA, {\it all} strong-[OI] sources and {\it all
but one} of the $\delta_K < 10\%$ objects have red gradients, the
exception being NGC 4150, a Young-TO for which, as discussed above,
$\delta_K$ is underestimated. On the other hand, galaxies with
significantly diluted $W_\lambda$ profiles, $\sim 90\%$ of which are
Young-TOs, have both blue and red gradients.  Of the 13 LLAGN with
$\delta_K > 10\%$, 8 have blue gradients and 5 have red gradients, but
note that in several of these objects the colour gradient is
negligible, with $|\delta_V| < 0.1$ mag.

%%%FIG%%%FIG%%%FIG%%%FIG%%%FIG%%%FIG%%%FIG%%%FIG%%%FIG%%%FIG%%%FIG%%%
\begin{figure*}
\psfig{file=Fig_dAV5313.eps2,width=14.2cm,bbllx=70,bblly=420,bburx=520,bbury=710}
\caption{Colour gradients, as given by the differential extinction
$\delta_V$ implied by the 5313/4020 colour, plotted against (a) the
dilution in the K line, (b) the nuclear equivalent width of the K
line, and (c) [OI]/H$\alpha$. Nuclei which are redder (bluer) than the
off-nuclear spectra have positive (negative) $\delta_V$.  Symbols as
in figure \ref{fig:dil_X_EW}.}
\label{fig:dAV5313}
\end{figure*}                                           
%%%FIG%%%FIG%%%FIG%%%FIG%%%FIG%%%FIG%%%FIG%%%FIG%%%FIG%%%FIG%%%FIG%%%

Because colours {\it per se} do not disentangle intrinsic stellar
population properties from extinction, these estimates of $\delta_V$
can only be interpreted as actual spatial variations in dust content
in the absence of stellar population variations. This is a reasonable
assumption for objects with relatively flat $W_\lambda$ profiles (and
thus spatially uniform stellar populations), which, as shown above,
are essentially all Old-LINERs and Old-TOs. The red-gradients observed
in these objects can thus be safely attributed to extinction
gradients.  Note, however, that most objects with $\delta_K \sim 0$
cluster around values of $\delta_V$ of 0.1--0.4 mag, indicating that
extinction gradients tend to be small.

The assumption of spatially uniform stellar populations breaks down
for galaxies with diluted or peaked $W_\lambda$ profiles. In the
latter case, one expects $\delta_V$ to overestimate the extinction
gradient, as the reference off-nuclear extractions sample a younger
population than that present in the nucleus. This is clearly the case
of NGC 6951, for which we obtain $\delta_V = 1.2$ mag, the largest
value in the whole sample. This effect is responsible for at least
part of the trend of increasing of $\delta_V$ as $\delta_K$ becomes
more negative (figure \ref{fig:dAV5313}a). Conversely, when the
nucleus contains a younger (and thus intrinsically bluer) population
than off-nuclear positions, the resulting $\delta_V$ should be
regarded as a {\it lower limit} to the actual variation in $A_V$.
From \S\ref{sec:dilution} and figure \ref{fig:dilutions_x_lambda} we
know that in galaxies with diluted $W_\lambda$ profiles the diluting
source is intrinsically bluer than off-nuclear spectra, which should
lead to blue gradients.  While some of these galaxies indeed have blue
gradients, most (9/13) have negligible or slightly red gradients, {\it
which can only be understood in terms of a higher dust content in the
nucleus}. Hence, contrary to the first impression derived from the
relative rarity of centrally peaked $C_{5313}$ profiles among these
sources, extinction gradients seem to be a common feature of
Young-TOs.

In summary, this empirical analysis shows that extinction gradients
are present in LLAGN of all kinds.  In Old-LINERs and Old-TOs, which
have spatially uniform stellar populations, these gradients are not
huge, with $\delta_V$ typically smaller than 0.5 mag. Young-TOs, with
their diluted $W_\lambda$ profiles, also have extinction gradients,
but a quantitative assessment of their magnitude requires a more
elaborate analysis, which we present in \S\ref{sec:extinction}.

\subsection{Slit Profiles}

\label{sec:SlitProfiles}

The central intermediate age population which dilutes the equivalent
widths of metal lines must cause an excess of flux with respect to the
smooth surface brightness profile from the bulge of the host galaxy.
Galaxies containing this extra central source should thus have sharper
brightness profiles than those with more uniform stellar populations.

\begin{figure*}
\psfig{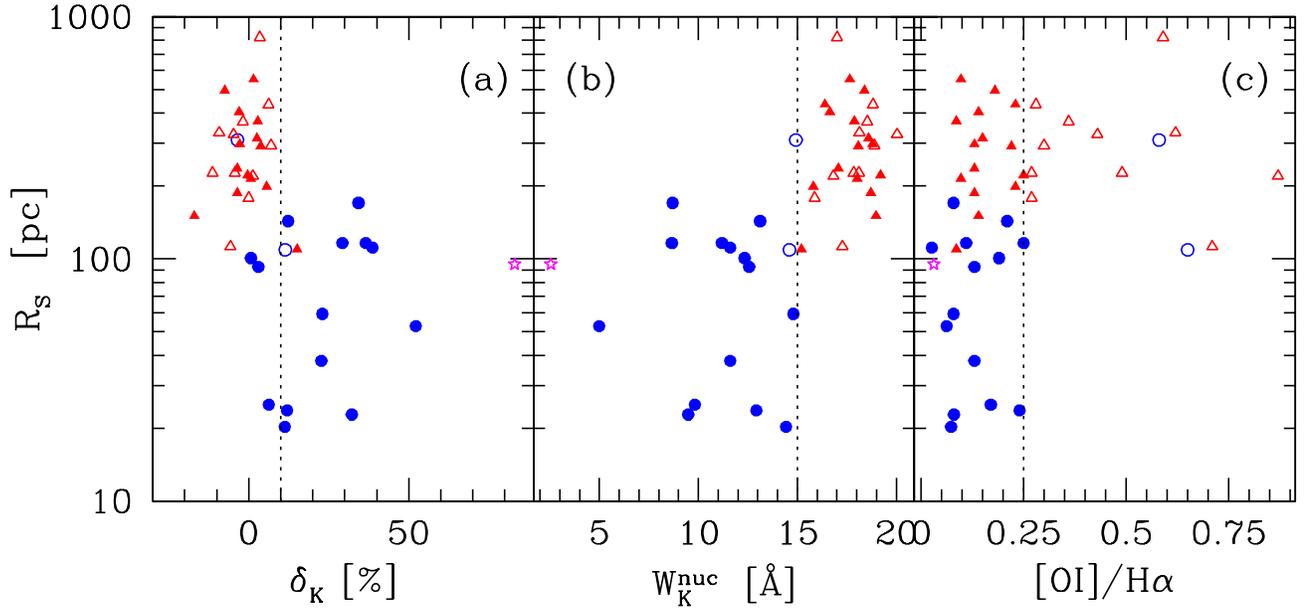}
\caption{Half Width at Half Maximum of the flux distribution along the
slit plotted against (a) the dilution in the K line, (b) the nuclear
equivalent width of the K line, and (c) [OI]/H$\alpha$.  Symbols as in
figure \ref{fig:dil_X_EW}.}
\label{fig:HWHMtot}
\end{figure*}

In order to verify this prediction we have measured the Half Width at
Half Maximum (HWHM) of the slit profiles, denoted by $R_S$.  The
results are listed in the last two columns of Table \ref{tab:dilution}
in angular and linear scales, and graphically illustrated in figure
\ref{fig:HWHMtot}. The plot confirms that the most compact slit
profiles occur among sources with diluted $W_K$ profiles.  By
extension of the relations between $\delta_K$, $W_K^{\rm nuc}$ and
[OI]/H$\alpha$, one expects these compact sources to be mostly
Young-TOs, as confirmed in figures \ref{fig:HWHMtot}b and c.

The slit profiles of Young-TOs suggests characteristic sizes of
50--100 pc for their central intermediate age population. This rough
estimate suffers from two caveats. First, it is based on the total
flux profile, which includes the bulge light. This issue is addressed
in \S\ref{sec:Analysis} below. Second, in several cases $R_S$
corresponds to angular sizes of $1^{\prime\prime}$ or less (Table
\ref{tab:dilution}), in which case seeing starts to dominate size
estimates. In fact, the comparison of galaxy and stellar profiles in
the bottom panels of figures
\ref{fig:grad_NGC0315}--\ref{fig:grad_NGC3367} shows that while Old
LLAGN have spatially resolved profiles, in Young TOs the inner $S(r)$
profile is only marginally broader than the seeing disk, so $R_S$
should be regarded as an {\it upper limit} for these objects. A more
refined study of the inner morphology of LLAGN based on high
resolution imaging is underway (Gonz\'alez Delgado \etp, in
preparation).

%%%FIG%%%FIG%%%FIG%%%FIG%%%FIG%%%FIG%%%FIG%%%FIG%%%FIG%%%FIG%%%FIG%%%
%%%FIG%%%FIG%%%FIG%%%FIG%%%FIG%%%FIG%%%FIG%%%FIG%%%FIG%%%FIG%%%FIG%%%

%%%SEC%%%SEC%%%SEC%%%SEC%%%SEC%%%SEC%%%SEC%%%SEC%%%SEC%%%SEC%%%SEC%%%

%%%SEC%%%SEC%%%SEC%%%SEC%%%SEC%%%SEC%%%SEC%%%SEC%%%SEC%%%SEC%%%SEC%%%
\section{Analysis and Discussion}

\label{sec:Analysis}

Our spatially resolved spectra of LLAGN show that significant stellar
population gradients occur almost exclusively in Young-TOs. These
gradients are caused mostly by an intermediate age population (0.1--1
Gyr), although in a few cases a $< 10$ Myr nuclear starburst is also
present (Papers I and II). The contribution of these stars to the
total spectrum increases towards the nucleus, causing the radial
dilution of metallic features. For consistence of notation, we
hereafter denote this population the {\it ``Central Young
Population''} (CYP), where ``young'' means $\la 1$ Gyr-old.

In this section we present estimates of the physical size, luminosity
and extinction of the CYPs in Young-TOs.  These estimates require
separating the light from the CYP from that of older stars from the
host's bulge, which in turn requires a more elaborated analysis than
the eminently empirical description of gradients presented in the
previous section. Two methods were developed with this purpose.  We
close this section with a discussion on what these CYPs looked like in
the past and what they might evolve to.

\subsection{Fits of the Equivalent Width profiles}

\label{sec:Wprof_fits}

A rough estimate of the size of the region responsible for the radial
dilution of metal lines may be obtained by evaluating at which
distance from the nucleus $W_K(r)$ crosses the dividing line at $W_K =
15$ \AA, which characterizes the transition from ``Young'' to ``Old''
stellar populations in our simple classification scheme. This is not
always possible, either because $W_K$ sometimes does not raise above
this threshold in the whole region analyzed (eg., NGC 4569, figure
\ref{fig:grad_NGC4569}) or because of asymmetries or oscillations in
the $W_K$ profile (eg., NGC 5678, figure \ref{fig:grad_NGC5678}).
For the objects where this analysis was possible, we estimate radii
between $\sim 100$ and 300 pc.

A more formal estimate may be obtained fitting the $W_\lambda$
profiles.  A two-components model was build for this purpose. We
assume that $W_\lambda(r)$ results from the superposition of a
``background'' component with a $W_\lambda(r) = W_\lambda(\infty)$
flat profile and a diluting component with negligible $W_\lambda$, and
whose fractional contribution $f(r)$ to the total continuum at
wavelength $\lambda$ and position $r$ follows a bell-shape radial
distribution. The resulting model is expressed by

\begin{equation}
\label{eq:wprof_fit}
W_\lambda(r) = W_\lambda(\infty)
               \left[ 1 - f(r) \right]
             = W_\lambda(\infty)
               \left[ 1 - \frac{\Delta_\lambda}{1 + (r/a_W)^2} \right]
\end{equation}

\noindent where $\Delta_\lambda = [W_\lambda(\infty) - W_\lambda(0)] /
W_\lambda(\infty)$ and $W_\lambda(\infty)$ are the analytical
equivalents of $\delta_\lambda$ and $W_\lambda^{\rm off}$ respectively
(see equation \ref{eq:dilution}). $a_W$ is the HWHM of the $f(r)$
profile, a size scale which should not be confused with the HWHM of
surface brightness profile associated with the central diluting
component.  The latter quantity, which we denote by $R_W$, must be
evaluated from the product of $S(r)$ and $f(r)$.

\begin{figure*}
\psfig{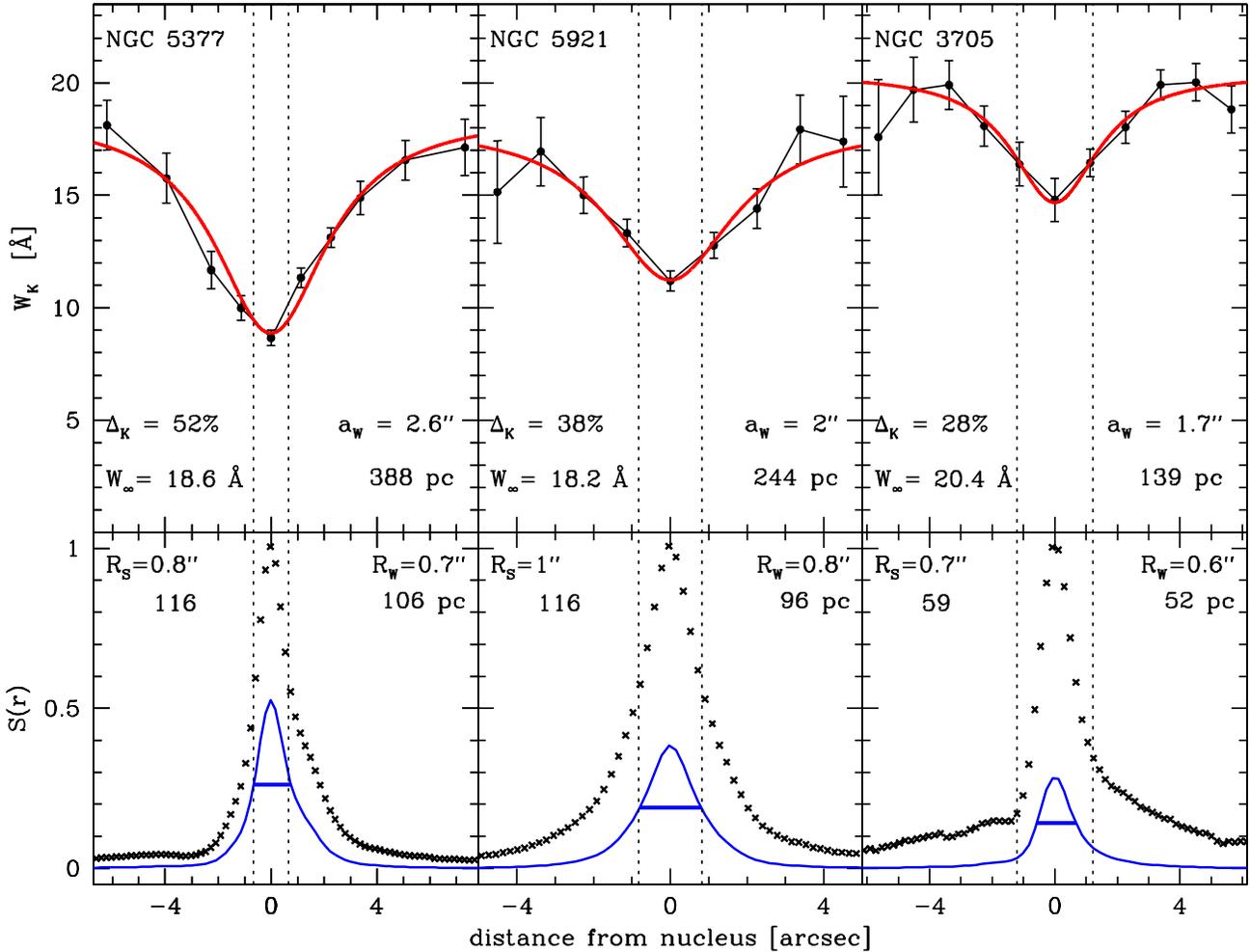}
\caption{{\it Top:} Examples of the $W_K(r)$ fits. {\it Bottom:}
Crosses show the normalized slit brightness profile. The solid blue
line shows the brightness profile of the diluting component inferred
from the $W_K(r)$ fits. Labels indicate the HWHM of the total
brightness profile ($R_S$) and the HWHM of the diluting source
($R_W$), also indicated by the thick horizontal line-segment.
Vertical dotted lines mark projected distances of $\pm 100$ pc from
the nucleus.}
\label{fig:WK_fits_examples}
\end{figure*}

We have fitted this model to the $W_K$ profiles of 15 LLAGN: NGC 404,
NGC 4150 plus the 13 LLAGN with $\delta_K > 10\%$ (ie, those with
significant dilution). The results are reported in Table
\ref{tab:WKfits_Results}.  The fits are generally good, as illustrated
in figure \ref{fig:WK_fits_examples}. The dilution factors obtained
from the fits are larger than the ones measured through equation
(\ref{eq:dilution}), with $\Delta_K \sim 1.3 \delta_K$ typically.
This happens because in most cases our operational definition of
$W_K^{\rm off}$ includes part of the rising portion of the $W_K(r)$
curve, while equation (\ref{eq:wprof_fit}) fits an asymptotic
value. Interestingly, we find $\Delta_K = 23$ and 29\% for NGC 404 and
NGC 4150 respectively, two Young-TOs for which $\delta_K$ fails to
detect significant dilution but, according to the qualitative
considerations in \S\ref{sec:dilution}, should be included in the list
of sources with diluted $W_K$ profiles.

%%%FIG%%%FIG%%%FIG%%%FIG%%%FIG%%%FIG%%%FIG%%%FIG%%%FIG%%%FIG%%%FIG%%%
%%%FIG%%%FIG%%%FIG%%%FIG%%%FIG%%%FIG%%%FIG%%%FIG%%%FIG%%%FIG%%%FIG%%%

%%%TAB%%%TAB%%%TAB%%%TAB%%%TAB%%%TAB%%%TAB%%%TAB%%%TAB%%%TAB%%%TAB%%%
\begin{table*}
%\tiny
\begin{centering}
\begin{tabular}{lrrrrrr}
\multicolumn{7}{c}{Results of the $W_K(r)$ fits}\\ \hline
NGC                        &
$W_K(\infty)$ [\AA]        &
$\Delta_K$ [\%]            &
$a_W$ [$^{\prime\prime}$]  &
$a_W$ [pc]                 &
$R_W$ [$^{\prime\prime}$]  &
$R_W$ [pc]                 \\ 
(1) &
(2) &
(3) &
(4) &
(5) &
(6) &
(7) \\ \hline
 %# Gal     wK_bck           dilK[%]       a["]           a[pc]            R["]           R[pc]
0404 & 12.2 $\pm$  0.4 &   23 $\pm$  4 & 3.7 $\pm$ 0.5 &   43 $\pm$    6 & 1.7 $\pm$ 0.1 &   20 $\pm$    1 \cr
0718 & 17.9 $\pm$  1.4 &   28 $\pm$  6 & 3.6 $\pm$ 0.6 &  374 $\pm$   65 & 1.2 $\pm$ 0.1 &  126 $\pm$    5 \cr
0772 & 19.9 $\pm$  0.3 &   42 $\pm$  3 & 1.1 $\pm$ 0.1 &  172 $\pm$   20 & 0.5 $\pm$ 0.1 &   86 $\pm$    3 \cr
3245 & 18.2 $\pm$  0.3 &   17 $\pm$  2 & 1.0 $\pm$ 0.3 &  107 $\pm$   29 & 0.6 $\pm$ 0.1 &   67 $\pm$    8 \cr
3627 & 15.5 $\pm$  0.3 &   25 $\pm$  3 & 1.4 $\pm$ 0.3 &   44 $\pm$    9 & 0.8 $\pm$ 0.1 &   25 $\pm$    2 \cr
3705 & 20.4 $\pm$  0.7 &   28 $\pm$  4 & 1.7 $\pm$ 0.5 &  139 $\pm$   39 & 0.6 $\pm$ 0.1 &   52 $\pm$    3 \cr
4150 & 15.7 $\pm$  1.3 &   29 $\pm$  7 & 4.2 $\pm$ 0.7 &  199 $\pm$   32 & 1.8 $\pm$ 0.1 &   82 $\pm$    3 \cr
4569 & 22.6 $\pm$  2.4 &   78 $\pm$  2 & 4.5 $\pm$ 0.5 &  365 $\pm$   37 & 0.6 $\pm$ 0.1 &   52 $\pm$    3 \cr
4736 & 15.6 $\pm$  0.2 &   17 $\pm$  2 & 2.3 $\pm$ 0.6 &   47 $\pm$   12 & 0.9 $\pm$ 0.1 &   19 $\pm$    1 \cr
4826 & 17.0 $\pm$  0.6 &   15 $\pm$  4 & 2.0 $\pm$ 0.7 &   39 $\pm$   14 & 0.8 $\pm$ 0.1 &   17 $\pm$    2 \cr
5005 & 16.5 $\pm$  0.3 &   15 $\pm$  3 & 1.7 $\pm$ 0.2 &  172 $\pm$   24 & 0.8 $\pm$ 0.1 &   85 $\pm$    4 \cr
5377 & 18.6 $\pm$  1.0 &   52 $\pm$  2 & 2.6 $\pm$ 0.4 &  388 $\pm$   59 & 0.7 $\pm$ 0.1 &  106 $\pm$    2 \cr
5678 & 13.2 $\pm$  0.7 &   33 $\pm$  4 & 1.1 $\pm$ 0.7 &  190 $\pm$  115 & 0.6 $\pm$ 0.2 &  110 $\pm$   34 \cr
5921 & 18.2 $\pm$  1.7 &   38 $\pm$  6 & 2.0 $\pm$ 0.5 &  244 $\pm$   63 & 0.8 $\pm$ 0.1 &   96 $\pm$    7 \cr
6503 & 14.4 $\pm$  0.6 &   34 $\pm$  4 & 1.1 $\pm$ 0.4 &   31 $\pm$   13 & 0.6 $\pm$ 0.1 &   17 $\pm$    3 \cr \hline
%# Output of fit_wprof.for (results for K line)
%# Cid@Lynx - 02/Feb/2004
%# (OBS: +/- 0 error in R[``] transformed to +/- 0.1 ...) Cid - 14/June/2004
\end{tabular}
\end{centering}
\caption{Results of the $W_K(r)$ fits for LLAGN with
diluted $W_K$ profiles.}
\label{tab:WKfits_Results}
\end{table*}
%%%TAB%%%TAB%%%TAB%%%TAB%%%TAB%%%TAB%%%TAB%%%TAB%%%TAB%%%TAB%%%TAB%%%

The values of $a_W$ range from $\sim 30$ to 400 pc, with a median of
172 pc, in agreement with the cruder estimates based on the size of
the $W_K < 15$ \AA\ region. These values are larger than the HWHM of
$S(r)$, which spans the $R_S = 20$--170 pc range, with a median of 93
pc for this subset of galaxies (Table \ref{tab:dilution}). In other
words, $f(r)$ is broader than $S(r)$.  Therefore, in practice the HWHM
of the light distribution associated with the CYP is dictated more by
the slit profile than by the $f(r)$ deduced from the $W_K(r)$
fits. The $S \times f$ profiles yield $R_W = 17$ to 126 pc (median =
67 pc), just slightly smaller than $R_S$.  Hence, although it is clear
that these CYPs often extend to more than 100 pc from the centre (as
demonstrated by the detection of HOBLs well outside the nucleus; eg.,
figures \ref{fig:offnuc_spectra_examples1} and
\ref{fig:offnuc_spectra_examples2}), most of their light is
concentrated within $r \la 100$ pc. The relatively little light from
the outer ($r > R_W$) parts of this distribution is enough to compete
with the flux from the host's bulge, producing diluted $W_\lambda$
profiles on scales $a_W$ substantially larger than $R_W$.

It is important to emphasize that in angular units, the median $R_W$
corresponds to just $0.8^{\prime\prime}$. Hence, although we are able
to resolve the wings of the light profile of CYPs, seeing prevents us
from adequately sampling their core. Our estimates of $R_W$ should
thus be regarded as {\it upper limits} to the actual CYP
radius. Indeed, high resolution images of a few Young-TOs reveal
structures on scales smaller than the ones we are able to trace with
our $\sim 1^{\prime\prime}$ resolution. For example, NGC 4569 is known
to have a very strong and compact nuclear source. Maoz \etal (1996)
found, based on an HST/FOC image at 2300 \AA, that the emission of
this galaxy is composed of a bright unresolved nuclear point source
and some faint extended emission $0.65^{\prime\prime}$ south of the
nucleus. Similar observations by Barth \etal (1998), done with
HST/WFPC2 at 2200 \AA, find that the nucleus is slightly resolved
along PA $= 20^{\circ}$, with a dimension of $0.16^{\prime\prime}
\times 0.11^{\prime\prime}$.

\subsection{Template decomposition}

\label{sec:TemplateDecomposition}

\subsubsection{Method}

\label{sec:TemplateDecompositionMethod}

An alternative and more complete way to analyze gradients in stellar
populations is to model each extraction in terms of a superposition of
spectra of well understood stellar populations. This can be achieved
by means of the empirical starlight modeling scheme introduced in
Paper I. The method consists of fitting a given spectrum with a
combination of five non-active galaxies from our comparison sample,
whose spectra represent stellar population classes $\eta = Y$ (NGC
3367), $I$ (NGC 205), $I/O$ and $O$ (NGC 221, NGC 1023 and NGC 2950).
The code outputs the fractional contribution of these components
to the flux at 4020 \AA, expressed as a {\it population vector} ${\bf
x} = (x_Y,x_I,x_O)$, where the $\eta = I/O$ and $O$ components are
grouped in $x_O$ for conciseness.  The code also fits the extinction
$A_V$, modeled as due to an uniform dust screen with $A_V$ up to 4
mag.  Regions around emission lines are masked out in the comparison
of model and observed spectra. Paper I shows that this method provides
excellent fits to the spectra.  Unlike in Papers I and II, we have
dereddened the template galaxies by their {\it intrinsic} extinction
derived by method described in Cid Fernandes \etal (2004b). Only NGC
3367 and NGC 205 are found to have significant extinction, both with
$A_V \sim 0.9$ mag. These corrections were applied because of our
interest in estimating the extinction and its radial variations in
LLAGN.

We have applied this method to all nuclear and off-nuclear spectra
analyzed in this paper, thereby producing stellar population and
extinction profiles. The spectral fits are of similar quality to those
exemplified in Paper I. The median fractional difference between model
and observed spectra for all extractions is 4.5\%, which is acceptable
considering a median noise-to-signal ratio of 6\% at 4000 \AA\ and 3\%
at 4800 \AA.

%%%FIG%%%FIG%%%FIG%%%FIG%%%FIG%%%FIG%%%FIG%%%FIG%%%FIG%%%FIG%%%FIG%%%
\begin{figure*}
\psfig{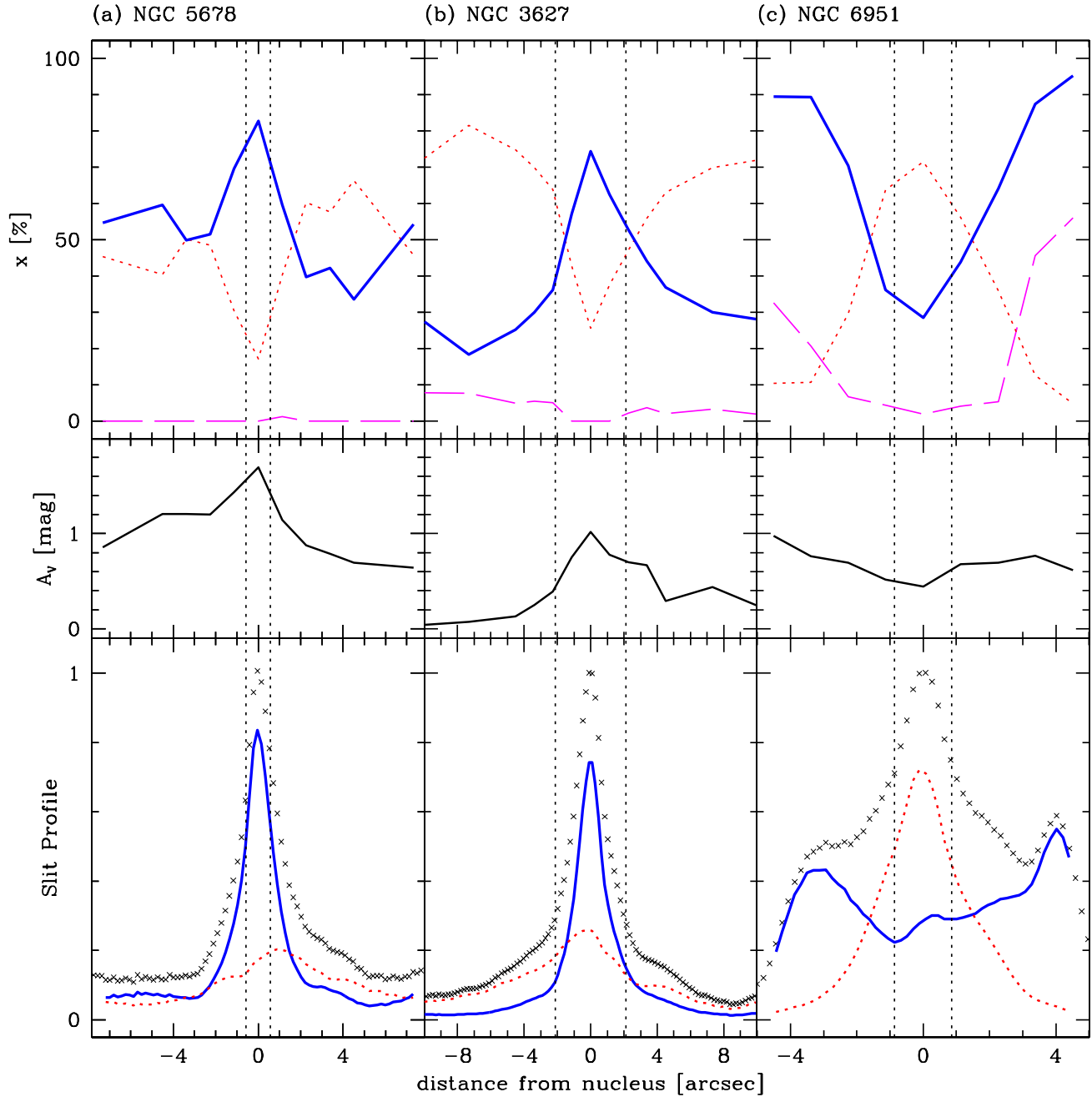}
%\resizebox{\textwidth}{!}{\includegraphics{Fig_3TFPexamples.eps2}}
\caption{Examples of the results of the spectral fit of the spatially
resolved spectra of four LLAGN using a base of template galaxies. {\it
Top:} Radial profile of the population vector. Old stellar populations
($x_{I/O} + x_O$) are represented by dotted (red) lines, young plus
intermediate and age populations ($x_{Y+I} = x_Y + x_I$) by a solid
(blue) line and young starbursts ($x_Y$) by a dashed line. {\it
Middle:} Radial profile of the extinction $A_V(r)$. {\it Bottom:}
Crosses show the observed surface brightness profile along the slit,
normalized to its peak value. The solid (blue) and dotted (red) lines
represent the profiles associated with the young + intermediate and
old components respectively. Vertical dotted lines indicate projected
distances of $\pm 100$ pc from the nucleus.}
\label{fig:3TFP_examples}
\end{figure*}

\begin{figure*}
\psfig{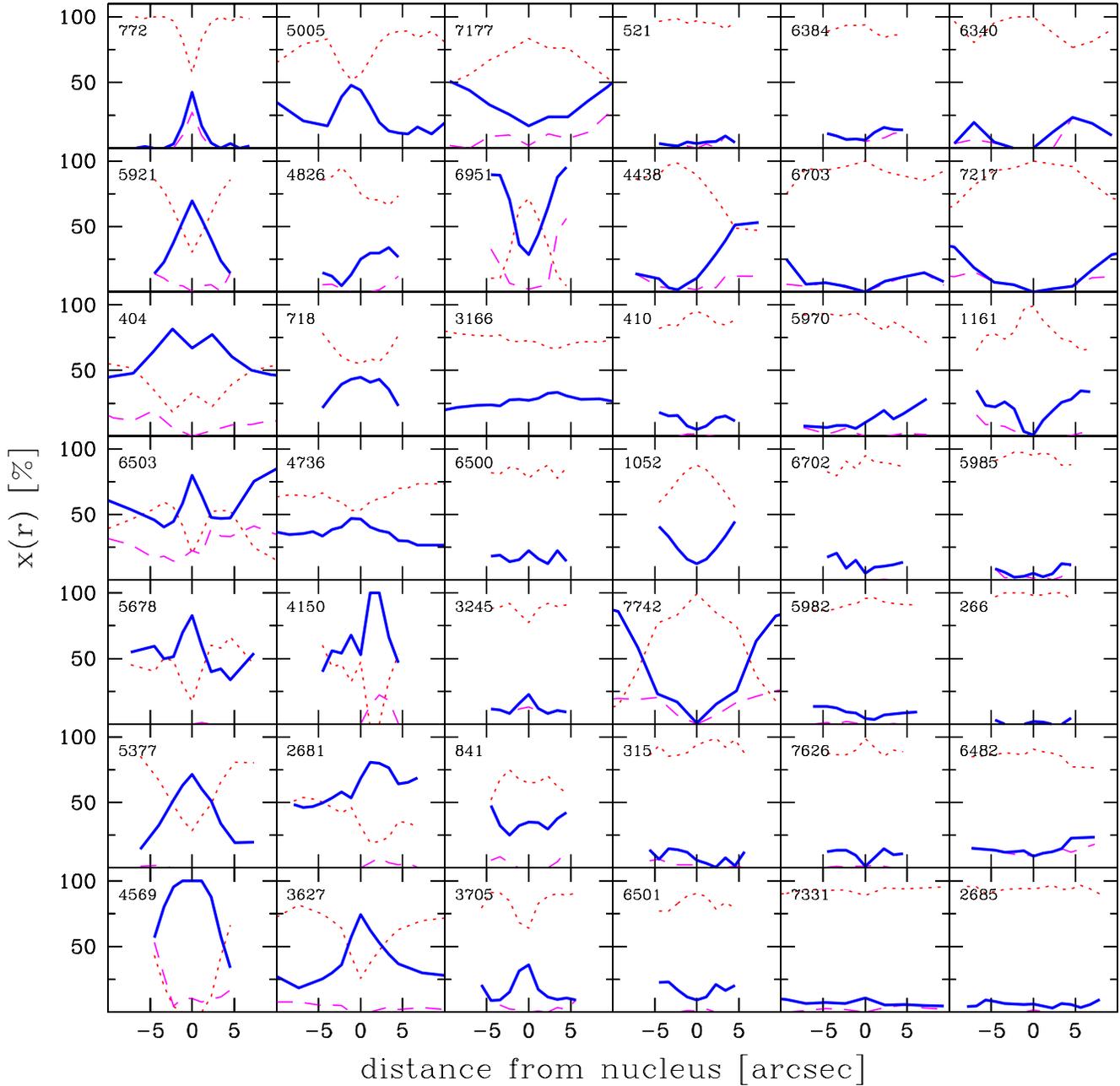}
%\resizebox{\textwidth}{!}{\includegraphics{Fig_all_x_TFP.eps2}}
\caption{Results of the template decomposition for all 42 LLAGN. Plots
are ordered according to the value of $W_K^{\rm nuc}$, from small
values in the bottom-left to large values in the top right.  Dotted,
red lines correspond to $x_O$; solid blue lines correspond to
$x_{Y+I}$ and dashed lines to $x_Y$.}
\label{fig:all_x_TFP}
\end{figure*}
%%%FIG%%%FIG%%%FIG%%%FIG%%%FIG%%%FIG%%%FIG%%%FIG%%%FIG%%%FIG%%%FIG%%%

Examples of the resulting ${\bf x}(r)$ and $A_V(r)$ are illustrated in
figure \ref{fig:3TFP_examples}.  The population vector in these plots
is grouped into a predominantly old component, $x_O$, (dotted red
line) and a young + intermediate age component, $x_{Y+I} = x_Y + x_I$
(solid blue line), representing the combined strengths of the NGC 3367
and NGC 205-like components.  This coarse 2-components description of
the stellar population matches our Young/Old classification scheme.
The $x_{Y+I}(r)$ fraction, in particular, is used to map the CYP. We
further plot $x_Y(r)$ as a dashed line to illustrate that $x_{Y+I}$ is
actually dominated by the intermediate age population. As found in
Papers I and II, young starbursts are generally weak or absent in
LLAGN, although off-nuclear star-formation occurs in a few cases, as
in NGC 6951. This Old-TO provides a good example of the power of the
method (figure \ref{fig:3TFP_examples}c).  Its well known star-forming
ring (P\'erez \etal 2000), at $r \sim 4^{\prime\prime} \sim 500$ pc,
is nicely mapped by the $x_{Y+I}$ profile and its associated
brightness distribution (bottom panel), obtained from the
multiplication of $x_{Y+I}(r)$ by the slit profile $S(r)$. Notice also
the rise in extinction in the ring, the presence of an intermediate
age component throughout the observed region, particularly in the
ring, and the prevalence of an old, bulge-like population in
the nucleus, which accounts for over 70\% of the light.

Figure \ref{fig:all_x_TFP} shows the $\vec{x}(r)$ profiles for all 42
LLAGN in our sample. As in previous plots, galaxies are ordered from
bottom-right to top-left in an increasing sequence of $W_K^{\rm nuc}$.
The plot confirms that the spectral gradients identified in
\S\ref{sec:GradentsInStellarProperties} are indeed associated with a
centrally concentrated intermediate age stellar population, plus, in a
few cases, a young starburst (eg., NGC NGC 772).  This can be seen by
the peaked $x_{Y+I}$ profiles from NGC 4569 to NGC 6500 in figure
\ref{fig:all_x_TFP}. Conversely, the old stellar component, mapped by
$x_O(r)$ in these plots, bears a clear similarity in shape with the
$W_\lambda$ profiles: Galaxies with diluted lines have diluted
$x_O(r)$ profiles. Similarly, the spatial homogeneity of stellar
populations inferred from the flat $W_\lambda(r)$ in galaxies like NGC
266 and most others in the right half of figure \ref{fig:all_x_TFP} is
confirmed by equally flat $x_O$ profiles, while peaked $W_\lambda$
profiles map onto peaked $x_O$ profiles (eg, NGC 1161).

Hence, to first order, the ${\bf x}(r)$ profiles obtained from the
template decomposition merely map the $W_\lambda$ variations onto the
stellar population space spanned by our normal galaxy base. In fact,
this relation is so strong that the equation

\begin{equation}
\label{eq:xo_X_WK}
x_O  = (0.068 \pm 0.001) W_K[{\rm \AA}] - (0.35 \pm 0.02)
\end{equation}

\ni transforms $W_K$ into $x_O$ to within better than 0.1 rms for all
521 spectra. Plugging our $W_K^{\rm nuc} = 15$ \AA\ dividing line in
this equation we find that the transition from Young to Old stellar
population occurs around $x_O \sim 2/3$, or, equivalently, $x_{Y+I}
\sim 1/3$. We thus conclude that CYPs which account for $\la 1/3$ of
the optical light would not be recognized as such in our data. Indeed,
of the 15 LLAGN with CYPs detected through the radial dilution of
$W_K$ only 2 have $x_{Y+I} < 1/3$: NGC 4826 ($W_K^{\rm nuc} = 14.4$
\AA\ and $x_{Y+I} = 0.25$) and NGC 3245 ($W_K^{\rm nuc} = 15.2$ \AA\
and $x_{Y+I} = 0.23$).

\subsubsection{Extinction profiles}

\label{sec:extinction}

Our empirical analysis of colour and equivalent width gradients in
\S\ref{sec:ColourGradients} indicates that extinction gradients are
generally small in Old-LLAGN, while for Young systems we could only
reach the qualitative conclusion that extinction variations must
occur.  A much more refined analysis is possible with the template
decomposition method, which produces quantitative estimates of both
gradients and absolute values of the extinction.

\begin{figure*}
\psfig{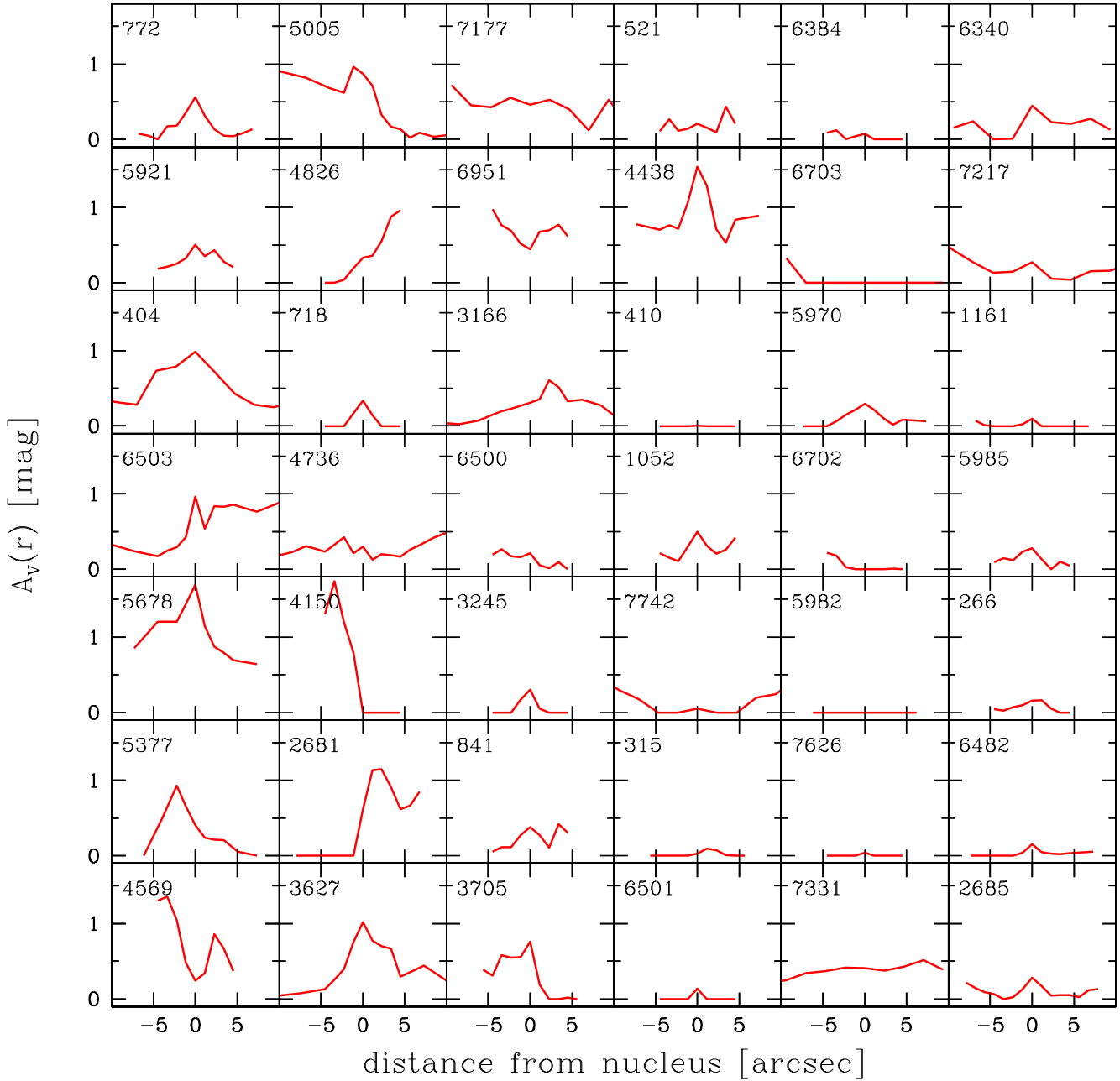}
%\resizebox{\textwidth}{!}{\includegraphics{Fig_all_AV_TFP.eps2}}
\caption{Extinction profiles obtained from the template decomposition
analysis. The ordering of the galaxies is as in Fig.\
\ref{fig:all_x_TFP}.}
\label{fig:all_x_AV_TFP}
\end{figure*}

The $A_V(r)$ profiles derived by this method are presented in figure
\ref{fig:all_x_AV_TFP} for our 42 LLAGN.  The first result which
strikes the eye in this plot is the obvious asymmetry between galaxies
in the left and right halves of the figure, which, given the ordering
according to $W_K^{\rm nuc}$, essentially correspond to Young and Old
systems respectively. The extinction profiles of Young-LLAGN are
substantially more complex than those of Old-LLAGN, which are often
approximately flat. In both cases, extinction gradients, when present,
are generally in the sense of producing centrally peaked $A_V$
profiles, indicating a higher concentration of dust in the central
regions. It is nevertheless clear that other types of dust
distribution exist, as in NGC 4150 and NGC 4826, whose asymmetric
$A_V(r)$ curves indicate the presence of off-nuclear dust-lanes.

A second and even more obvious result from figure
\ref{fig:all_x_AV_TFP} is that there is a clear offset in the absolute
values of $A_V$ between Young and Old systems. The statistics of $A_V$
reflect this difference. Averaging $A_V(r)$ over all extractions for
each galaxy, we obtain a median spatially-averaged extinction of
0.42 for our 16 Young-LLAGN, compared to 0.11 for the 26 Old-LLAGN. A
similar off-set is found considering only the nuclear extractions,
which have median $A_V(0) = 0.62$ and 0.21, respectively.
Young-LLAGN, $\sim 90\%$ of which are Young-TOs, are therefore $\sim
3$ times dustier than Old-LLAGN.  The clearest exception to this
strong correlation is NGC 4438. The high concentration of dust
inferred from $A_V$ profile of this Old-LINER is associated with the
pronounced nuclear dust lane seen in HST images (Kenney \& Yale 2002).

The Balmer decrement measurements of HFS97 lend further support to
interpretation that Young-TOs have a higher dust content than other
LLAGN. Using their tabulated values for objects in our sample, we find
a median H$\alpha/$H$\beta$ of 4.6 for Young-TOs and 3.1 for other
LLAGN. We can extend this analysis to the whole HFS97 sample using
their measurements of the G-band equivalent width and classifying
LLAGN into Young or Old adopting a $W$(G-band) = 4 \AA\ dividing line,
which is roughly equivalent to our Young/Old division at $W_K = 15$
\AA\ (Paper I). The 27 Young-TOs in this larger sample have a median
H$\alpha/$H$\beta = 4.5$, while for the other 116 LLAGN this ratio is
3.2.

We thus conclude that all evidence points towards a scenario where
Young-TOs are the dustier members of the LLAGN family.

%%%FIG%%%FIG%%%FIG%%%FIG%%%FIG%%%FIG%%%FIG%%%FIG%%%FIG%%%FIG%%%FIG%%%
%%%FIG%%%FIG%%%FIG%%%FIG%%%FIG%%%FIG%%%FIG%%%FIG%%%FIG%%%FIG%%%FIG%%%

\subsubsection{Sizes and luminosities of the CYPs}

\label{sec:SizesAndLuminosities_TemplateDecomposition}

The population vector derived through the template decomposition
analysis may be combined with the slit-profiles to produce
one-dimensional surface-brightness profiles of the different stellar
populations in our galaxies, as illustrated in the bottom panels of
figure \ref{fig:3TFP_examples}. In what follows we use this method to
estimate sizes and luminosities of the CYPs, represented by the
$S_{CYP}(r) = S(r) \times x_{Y+I}(r)$ profile.  This method differs
from the one in \S\ref{sec:Wprof_fits} in two aspects: (1) instead of
assuming a functional form for the light fraction associated to the
CYP we derive this fraction empirically from the template
decomposition; and (2) all the spectrum is used, as opposed to a
single equivalent width.

\begin{figure*}
\psfig{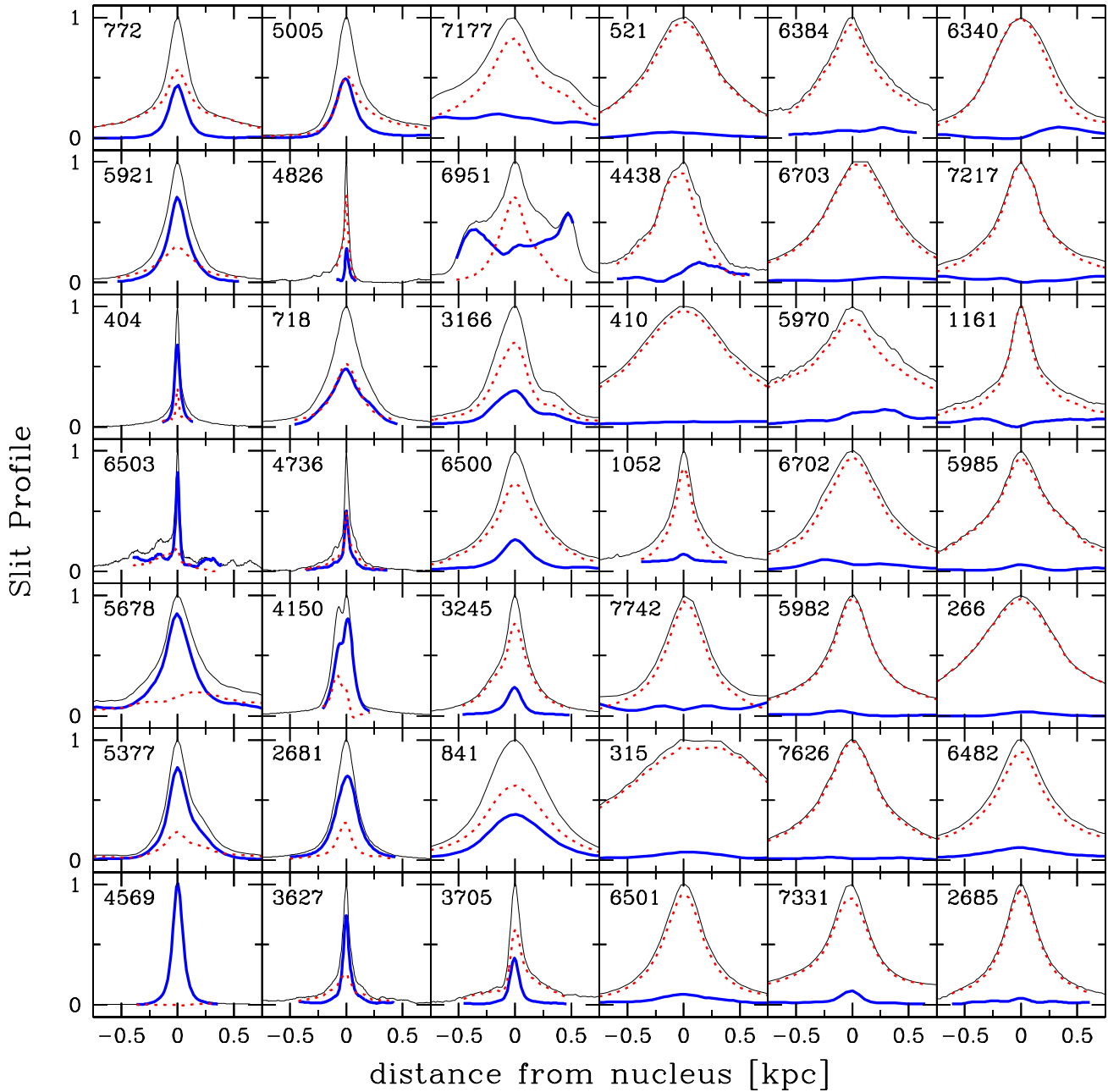}
%\resizebox{\textwidth}{!}{\includegraphics{Fig_all_SBprofs_TFP_kpc.eps2}}
\caption{Normalized slit brightness profiles (thin black line),
decomposed into young (thick solid blue line) and old components
(dotted red). The ordering of the galaxies is as in Fig.\
\ref{fig:all_x_TFP}.}
\label{fig:all_x_SBprof_TFP_kpc}
\end{figure*}

Figure \ref{fig:all_x_SBprof_TFP_kpc} shows the total slit profile
$S(r)$ (thin black line), and its decomposition into Young (thick blue
line) and Old (dotted red) components for our 42 LLAGN. The plot shows
that the young components dominate the light in the inner $\sim 100$
pc from NGC 4569 up to NGC 718 ($W_K^{\rm nuc} = 13.1$ \AA), becoming
fainter than the inner old population as $W_K^{\rm nuc}$ increases,
until it eventually ``vanishes'' from NGC 7177 onwards ($W_K^{\rm nuc}
> 16.6$ \AA). Note that, unlike all other profiles in this paper,
figure \ref{fig:all_x_SBprof_TFP_kpc} uses a linear scale for $r$,
which emphasizes the compactness of the CYPs in Young-TOs.

We estimate the radius of the CYPs from the HWHM of the $S_{CYP}(r)$
profiles.  Table \ref{tab:TempDecomp_Results} presents our results.
As for the $W_K(r)$ fits, we obtain $x_{Y+I}$ profiles which are
broader than $S(r)$, so $R_{CYP}$ is close to $R_S$ (Table
\ref{tab:dilution}). The values of $R_{CYP}$ are in good agreement
with $R_W$ (Table \ref{tab:WKfits_Results}), which is the equivalent
CYP radius in the $W_K(r)$ fits. Again, these estimates should be
regarded as {\it upper limits} given that the angular sizes are
limited by our spatial resolution.

The luminosity associated with the CYPs was estimated integrating
$S_{CYP}(r)$ within $|r| < 5 R_S$. The integration is performed in
half-rings of area $\pi r dr$, ie., extrapolating our 1D profiles to
2D.  Table \ref{tab:TempDecomp_Results} lists both the total and CYP
luminosities. Numbers in between parentheses correspond to
luminosities corrected by intrinsic extinction using the modeled $A_V$
profiles. The resulting dereddened CYP luminosities at 4020 \AA\ range
from $L_{CYP} \sim 10^{3.3}$ to $10^{5.5}$ L$_\odot$/\AA, with a
median of $10^{4.3}$ L$_\odot$/\AA. Expressed in more
conventional units, this roughly corresponds to a range in B-band
absolute magnitudes\footnote{We use a $M_B \approx -2.5 \log L_{4020}
- 3.96$ conversion, for $L_{4020}$ in units of L$_\odot$/\AA, derived
from the Starburst99 models.} from $\sim -12.2$ to -17.7, with a
median $M_B = -14.7$. Given the uncertainties in absolute flux
calibration, extinction correction and extrapolation from 1D to 2D
profiles, these values should be taken as order of magnitude
estimates. Yet, they are precise enough for the general considerations
we present next.

%%%FIG%%%FIG%%%FIG%%%FIG%%%FIG%%%FIG%%%FIG%%%FIG%%%FIG%%%FIG%%%FIG%%%
%%%FIG%%%FIG%%%FIG%%%FIG%%%FIG%%%FIG%%%FIG%%%FIG%%%FIG%%%FIG%%%FIG%%%

%%%TAB%%%TAB%%%TAB%%%TAB%%%TAB%%%TAB%%%TAB%%%TAB%%%TAB%%%TAB%%%TAB%%%
\begin{table*}
\begin{centering}
\begin{tabular}{lcrrr}
\multicolumn{5}{c}{CYP Sizes \& Luminosities from Template
Decomposition}\\ \hline
NGC                           &
$R_{CYP}$ [$^{\prime\prime}$] &
$R_{CYP}$ [pc]                &
$\log L_{tot}$                &
$\log L_{CYP}$                \\ 
(1) &
(2) &
(3) &
(4) &
(5) \\ \hline
%# S03Table-file aR_YOU    R_YOU  logL_tot (dered)   logL_YOU (dered)
0404$^\star$  &   2.5  &    29  &  3.77  (4.08)    &  3.54  (3.87)    \cr
0718          &   1.3  &   133  &  5.14  (5.17)    &  4.59  (4.64)    \cr
0772          &   0.5  &    79  &  5.06  (5.21)    &  4.10  (4.32)    \cr
3245          &   0.7  &    74  &  5.15  (5.19)    &  4.18  (4.22)    \cr
3367          &   0.4  &    90  &  5.00  (5.35)    &  4.91  (5.28)    \cr
3627          &   0.9  &    30  &  4.31  (4.60)    &  3.93  (4.26)    \cr
3705          &   0.6  &    48  &  4.24  (4.44)    &  3.48  (3.72)    \cr
4150          &   2.1  &    98  &  4.53  (5.00)    &  4.15  (4.70)    \cr
4569          &   0.6  &    53  &  5.20  (5.53)    &  5.19  (5.50)    \cr
4736          &   1.0  &    21  &  4.64  (4.77)    &  4.20  (4.34)    \cr
4826          &   0.9  &    19  &  3.89  (4.05)    &  3.11  (3.36)    \cr
5005          &   0.9  &    98  &  5.15  (5.45)    &  4.55  (4.93)    \cr
5377          &   0.7  &   104  &  5.11  (5.36)    &  4.85  (5.10)    \cr
5678          &   0.8  &   135  &  4.89  (5.53)    &  4.60  (5.27)    \cr
5921          &   0.8  &    95  &  5.03  (5.21)    &  4.63  (4.83)    \cr
6503          &   0.7  &    19  &  3.30  (3.61)    &  3.01  (3.33)    \cr \hline
%# Output of DecompSBprof.for
%# Cols 2   (aR): HWHM in arcsec
%# Cols 3    (R): HWHM in pc
%# Cols 4-7  (L): log Luminosity [L_sun/A]
%# Cid@Lynx - 16/May/2004
\end{tabular}
\end{centering}
\caption{CYP size and luminosities estimates from the template
decomposition analysis. Columns 4 and 5 give the total and CYP
monochromatic luminosities at 4020 \AA\ integrated along the slit and
extrapolated to 2D, in units of L$_\odot\,$\AA$^{-1}$. Numbers in
between parentheses are the dereddened luminosities.  $^\star = $
Observed under non-photometric conditions.}
\label{tab:TempDecomp_Results}
\end{table*}
%%%TAB%%%TAB%%%TAB%%%TAB%%%TAB%%%TAB%%%TAB%%%TAB%%%TAB%%%TAB%%%TAB%%%

\subsection{Discussion: The past and future of Young-TOs}

\label{sec:discussion}

Naturally, the intermediate age stars which typify the CYPs of
Young-TOs have been younger in the past and will get older in the
future. Their current age and luminosity can be used, with the aid of
evolutionary synthesis models, to predict what these objects looked
like in their early days and what they will eventually become.

For simplicity, lets assume that CYPs formed in instantaneous bursts
$10^8$--$10^9$ yr ago. From the Starburst99 models of Leitherer \etal
(1999) one infers that these CYPs were $\sim$ {\it 10 to 100 times
more luminous} in the optical in their first Myrs of life. Since the
old stellar population has not changed substantially over this period,
the CYPs would be much easier to detect back then. The {\it weakest}
CYPs recognized as such in our sample (ie, those with $W_K^{\rm
nuc} \le 15$ \AA) presently account for $x_{Y+I} \sim 33\%$ of the
nuclear light at $\lambda4020$. Scaling their present luminosity by
factors of 10--100 would raise this fraction to 83--98\%, which shows
that they would completely outshine the bulge light, and the optical
continuum would be essentially identical to that of a starburst
galaxy. Recall, however, that Young-TOs are dusty, so these luminous
infant CYPs could be substantially reddened and thus powerful far-IR
sources, particularly if they had even more dust (and gas) in their
early phases.

The hot, massive stars in these early phase would have a large impact
in the ionizing photon field. The H$\alpha$ to $\lambda4020$ flux
ratio for young starbursts is of order 1000 \AA\ (Leitherer \etal
1999). Currently, CYPs have $L_{\lambda4020} \sim 10^{4.3}$
L$_\odot\,$\AA$^{-1}$ (Table~\ref{tab:TempDecomp_Results}), which
scaled back to $t = 0$ yields H$\alpha$ luminosities of order
$10^{42}$ erg$\,$s$^{-1}$, more than two orders of magnitude larger
than those currently observed in Young-TOs and LLAGN in general, which
range from $10^{38}$ to $10^{40}$ erg$\,$s$^{-1}$ (HFS97). In terms of
$L_{H\alpha}$, they would rank among powerful starburst nuclei and
Seyferts. Clearly, these objects would definitely {\it not} be
classified as ``Low Luminosity'' in their youth.  It is not clear
whether they would be classified as AGN either!  Unless the AGN too
was much brighter in the past, these objects would surely look like
starbursts

Simple stellar populations of ages between $t \sim 10^8$ and $10^9$ yr
have mass-to-light ratios at $\lambda$4020 of $\sim 500$ to 5000
M$_\odot\,$L$_\odot^{-1}\,$\AA\ in the solar metallicity models of
Bruzual \& Charlot (2003). For a median CYP luminosity of $10^{4.3}$
L$_\odot\,$\AA$^{-1}$ (Table \ref{tab:TempDecomp_Results}), this
implies CYP masses $M_{CYP} \sim 10^7$--$10^8$
M$_\odot$. Star-formation has either ceased long ago or proceeds at a
residual level in CYPs, otherwise they would look much younger. It is
thus reasonable to suppose that these stars formed over a period of
time whose length is a fraction of their current age. For
star-formation time scales of $10^7$--$10^8$ yr, the typical
star-formation rate was of order 1 M$_\odot\,$yr$^{-1}$. These are
clearly very rough estimates, but they serve to set the scale of the
CYP phenomenon.

The precursors of Young-TOs thus have to be luminous nuclei with
substantial amounts of star-formation and possibly a bright AGN
too. Another clue is that these precursors must be found in the
local-universe, since $t \le 10^9$ yr corresponds to $z < 0.1$ for any
reasonable cosmology.  Two plausible contenders for the progenitors of
Young-TOs are starburst nuclei and starburst + Seyfert 2 composites
like Mrk 477, Mrk 1210 and others (Heckman \etal 1997;
Storchi-Bergmann, Cid Fernandes \& Schmitt 1998; Gon\'zalez Delgado,
Heckman \& Leitherer 2001). Given the tendency of TOs to have later
Hubble types than LINERs and Seyferts (Ho, Filippenko \& Sargent
2003), it seems more attractive to link Young-TOs with starburst
nuclei. However, the substantial overlap in morphological properties
between TOs and both starburst and AGN hosts, coupled to indications
that starburst + Seyfert 2 composites have rather late type
morphologies for AGN (Storchi-Bergmann \etal 2001) prevents us from
drawing a firm conclusion at this stage.

Similar arguments can be used to sketch the future evolution of
Young-TOs. As the CYP fades, it will eventually cross the $x_{Y+I} =
1/3$ threshold below which we would not identify it anymore and the
system would be classified as Old
(\S\ref{sec:TemplateDecompositionMethod}). For instance, starting from
a current value of $x_{Y+I} = 2/3$, and assuming the old populations
does not change much, the CYP would cross the $x_{Y+I} = 1/3$ line
after it fades by a factor of 4. For an assumed age of 1 Gyr, this
would take $\sim 2$ Gyr to happen. In other words, the stellar
populations of Young-TOs will become indistinguishable from those of
Old-LLAGN in a few Gyrs.  Though it is tempting to link Young to
Old-TOs because of their identical emission line properties, as
pointed out in Paper II we cannot rule out the possibility that
[OI]/H$\alpha$ increases as the CYP fades, which would turn a Young-TO
into an Old-LINER. Note also that for either of these two evolutionary
connections to work Young-TOs must some how get rid of their excess
dust (\S\ref{sec:extinction}) in a few Gyr, either by converting it to
new stars or blowing it away.

Although much work remains to be done, these general considerations
illustrate how the careful dissection of stellar populations
properties can provide new and important pieces in the quest to solve
the puzzle of active galactic nuclei.  The evolutionary scenarios
sketched above will be examined more closely in forthcoming
communications.
%%%SEC%%%SEC%%%SEC%%%SEC%%%SEC%%%SEC%%%SEC%%%SEC%%%SEC%%%SEC%%%SEC%%%

%%%SEC%%%SEC%%%SEC%%%SEC%%%SEC%%%SEC%%%SEC%%%SEC%%%SEC%%%SEC%%%SEC%%%
\section{Conclusions}

\label{sec:Conclusions}

In this third paper in our series dedicated to the stellar populations
of LLAGN, we have investigated the radial variations of stellar
populations properties in a sample of 42 LINERs and TOs plus 5
non-active galaxies.  The analysis was based on high quality
3500--5500 \AA\ long-slit spectra covering angular regions of at least
$\sim 10^{\prime\prime}$ in diameter with a resolution of $\sim
1^{\prime\prime}$ (corresponding to $\sim 100$ pc).

The main result of Papers I and II was the identification of a
class of objects which stand apart from other LLAGN in having a strong
$10^{8-9}$ yr population. In terms of emission lines nearly all of
these nuclei have weak [OI]/H$\alpha$, hence their denomination as
``Young-TOs''.  Here we have shown that Young-TOs are also distinct
from other LLAGN in terms of the way stellar populations and dust are
spatially distributed.  This general conclusion was reached through
two distinct and complementary ways.

First, radial profiles of absorption line equivalent widths,
continuum colours and the total flux along the slit were used to trace
the spatial distribution of stellar populations. The results of this
empirical analysis can be summarized as follows.

\begin{enumerate}

\item We find that the $W_\lambda$ profiles are of essentially two
types: flat and diluted.  Flat profiles, which indicate spatially
uniform stellar populations, are more common, accounting for $\sim
60\%$ of the sample. They occur exclusively in galaxies dominated by
an old, bulge-like stellar population, regardless of the LINER/TO
emission line classification.

\item Diluted profiles, on the other hand, are produced by a central
``young'' population (CYP) dominated by stars of $10^8$--$10^9$ yr,
whose relatively blue continuum dilutes the $W_\lambda$'s of metal
lines with respect to their off-nuclear values.

\item Although concentrated in the nucleus, these CYPs are spatially
extended, reaching distances of up to 400 pc from the nucleus.

\item The relation between diluted profiles and nuclear stellar
population is clearly expressed by the $\sim$ one-to-one relation
between the radial dilution index $\delta_\lambda$ and the nuclear
$W_\lambda$ for the CaII K line: Virtually {\it all} sources with
$\delta_K > 10\%$ have $W_K^{\rm nuc} < 15$ \AA\ and vice-versa.  This
range of $W_K^{\rm nuc}$ corresponds exactly to our definition of
``Young'' stellar population, meaning populations of 1 Gyr or
less. 

\item Since these stars are found almost exclusively in objects with
[OI]/H$\alpha \le 0.25$ (Papers I and II), it follows that stellar
population gradients are typical of Young-TOs.  The fact that these
stars are located in their central regions, and not spread over the
whole galaxy, reinforces the suggestion that they are somehow
connected to the ionization of the nuclear gas.

\ls Second, a more detailed analysis of stellar population
gradients was achieved by means of a decomposition of each spectra in
terms of templates representative of very young ($\le 10^7$ yr),
intermediate age ($10^{8-9}$ yr) and old ($10^{10}$ yr) stellar
populations. This analysis shows that: \ls

\item The CYPs in Young-TOs account for at least $\sim 1/3$ of the
total flux at 4020 \AA. We confirm the finding of Papers I and II that
these populations are dominated by $10^8$--$10^9$ yr stars. Young
starbursts, even when present, make a small contribution to the
optical light.

\item Yet another property which distinguishes Young-TOs from other
members of the LLAGN family is dust content.  Young-TOs are $\sim 3$
times more extincted than Old-LINERs and Old-TOs. This finding is
confirmed using the HFS97 measurements of the H$\alpha$/H$\beta$
ratio.

\item Dust tends to be concentrated towards the nucleus, although
asymmetric extinction profiles are also common.

\item The radial flux distribution of CYPs have HWHM radii of $\sim
100$ pc or less. While their core is at best partly resolved in our
data, their outer regions are clearly resolved.

\item The 4020 \AA\ luminosities of the CYPs are within an order of
magnitude of $10^{4.3}$ L$_\odot\,$\AA$^{-1}$, implying B-band
absolute magnitudes of $\sim -15$ and masses of order $\sim
10^7$--$10^8$ M$_\odot$. This population was 10--100 times more
luminous in their formation epoch, at which time young massive stars
would have completely outshone the bulge light.  The active nucleus
would also be swamped by these young starbursts, unless it too was
brighter in the past.

\end{enumerate}

This investigation has unveiled several interesting connections
between stellar population, emission line properties, spatial
distribution and extinction, paving the road to a better understanding
of the physics of low luminosity AGN. Future papers in this series
will explore these and other connections in further detail.
%%%SEC%%%SEC%%%SEC%%%SEC%%%SEC%%%SEC%%%SEC%%%SEC%%%SEC%%%SEC%%%SEC%%%

\section*{ACKNOWLEDGMENTS}

RCF and TSB acknowledge the support from CNPq, PRONEX and Instituto do
Mil\^enio. RGD acknowledges support by Spanish Ministry of Science and
Technology (MCYT) through grant AYA-2001-3939-C03-01.  The data
presented here have been taken using ALFOSC, which is owned by the
Instituto de Astrof\'{\i}sica de Andaluc\'{\i}a (IAA) and operated at
the Nordic Optical Telescope under agreement between IAA and the
NBIfAFG of the Astronomical Observatory of Copenhagen. We are grateful
to the IAA director for the allocation of 5.5 nights of the ALFOSC
guaranteed time. Data were also taken at Kitt Peak National
Observatory, National Optical Astronomy Observatories, which are
operated by AURA, Inc., under a cooperative agreement with the
National Science Foundation. The National Radio Astronomy Observatory
is a facility of the National Science Foundation, operated under
cooperative agreement by Associated Universities, Inc. This research
made use of the NASA/IPAC Extragalactic Database (NED), which is
operated by the Jet Propulsion Laboratory, Caltech, under contract
with NASA.

\end{document}